\documentclass[12pt,preprint]{aastex}
\shorttitle{Tools for Dissecting Supernova Remnants}
\shortauthors{LOPEZ ET AL.}

\newcommand{\xmm}{\textit{XMM}}
\newcommand{\xmmn}{\textit{XMM-Newton}}

\begin{document}

\title{Tools for Dissecting Supernova Remnants Observed with Chandra: Methods and Application to the Galactic Remnant W49B}

\author{
L.~A. Lopez\altaffilmark{1,2}, E.~Ramirez-Ruiz\altaffilmark{1},
D.~A. Pooley\altaffilmark{3}, T. E. Jeltema\altaffilmark{4}
}
%%%
\altaffiltext{1}{Department of Astronomy and Astrophysics, University of California Santa Cruz, 159 Interdisciplinary Sciences Building, 1156 High Street, Santa Cruz, CA 95064, USA; lopez@astro.ucsc.edu.}
\altaffiltext{2}{National Science Foundation Graduate Research Fellow}
\altaffiltext{3}{Astronomy Department, University of Wisconsin, 4512 Sterling Hall, 475 North Charter St, Madison, WI 53706-1582}
\altaffiltext{4}{UCO/Lick Observatories}

\begin{abstract}

We introduce methods to quantify the X-ray morphologies of supernova remnants observed with the {\it Chandra} X-ray Telescope. These include a power-ratio technique to measure morphological asymmetries, correlation-length analysis to probe chemical segregation and distribution, and wavelet-transform analysis to quantify \hbox{X-ray} substructure. We demonstrate the utility and accuracy of these techniques on relevant synthetic data. Additionally, we show the methods' capabilities by applying them to the 55-ks {\it Chandra} ACIS observation of the galactic supernova remnant W49B. We analyze the images of prominent emission lines in W49B and use the results to discern physical properties. We find that the iron morphology is very distinct from the other elements: it is statistically more asymmetric, more segregated, and has 25\% larger emitting substructures than the lighter ions. Comparatively, the silicon, sulfur, argon, and calcium are well-mixed, more isotropic, and have smaller, equally-sized emitting substructures. Based on fits of \xmmn\ spectra in regions identified as iron rich and iron poor, we determine that the iron in W49B must have been anisotropically ejected. We measure the abundance ratios in many regions, and we find that large, local variations are persistent throughout the remnant. We compare the mean, global abundance ratios to those predicted by spherical and bipolar core-collapse explosions; the results are consistent with a bipolar origin from a $\sim$25 $M_{\sun}$ progenitor. We calculate the filling factor of iron from the volume of its emitting substructures, enabling more precise mass estimates than previous studies. Overall, this work is a first step toward rigorously describing the physical properties of supernova remnants for comparison within and between sources. 

\end{abstract}

\keywords{methods: data analysis --- nuclear reactions, nucleosynthesis, abundances --- supernova remnants --- techniques: image processing --- X-rays: ISM}

\section{Introduction}

Supernova remnants (SNRs) are a diverse class of objects that play an essential role in the dynamics of the interstellar medium (ISM) and that produce and distribute most of the metals in the Universe \citep{fp04}. Since its launch, the {\it Chandra} X-ray Telescope has observed over one-hundred SNRs in the Milky Way galaxy \citep{gr06}. The sub-arcsecond spatial resolution and spectro-imaging capabilities of {\it Chandra} have facilitated detailed studies of the metal-rich ejecta from supernova explosions as well as the interactions with their environments as they expand (see Weisskopf \& Hughes 2006 for a review). {\it Chandra} has revealed that X-ray spectral and spatial morphologies of SNRs are extremely complex. However, no systematic, quantitative methods for comparison within and between SNRs have been applied to develop a global picture of these varied sources. 

Rigorous mathematical techniques are necessary to maximally utilize these {\it Chandra} datasets. Toward this end, we apply well-estalished tools to characterize the X-ray morphologies of SNRs. We present three methods especially suited for application to {\it Chandra} Advanced CCD Imaging Spectrometer (ACIS) observations: a multipole power-ratio technique to probe morphological symmetry, correlation-length analysis (an adaptation of two-point correlation) to measure chemical segregation and distribution, and wavelet-transform analysis to quantify substructure. These methods have been used previously in other astronomical contexts, and we have extended and further developed them for SNR science. In $\S$2, we describe the methods in detail and demonstrate their application to synthetic data of extended sources. In $\S$3, we apply these techniques to the 55-ks {\it Chandra} ACIS observation of W49B, a SNR with a particularly complex spatial distribution. In $\S$4, we discuss how the results can be used to discern between the heating and explosion properties of the source. We present our concluding remarks in $\S$5. 

\subsection{W49B Background}

W49B (G43.3$-$0.2) is a principal example of a galactic SNR with complex X-ray morphology. It is the brightest ejecta-dominated SNR in X-rays ($L_{X} \sim 10^{38}$ erg s$^{-1}$; Immler \& Kuntz 2005), and it has one of the highest surface brightnesses at 1 GHz of all galactic sources \citep{mr94}. Radio maps of the source show W49B has a prominent shell structure $\approx$4\arcmin\ in diameter, yet it has a center-filled morphology in the X-rays. HI absorption measurements indicate the remnant is 8$\pm$2 kpc away \citep{mr94}, and due to its distance through the Galactic plane, no optical emission is detected from the source. A recent {\it Spitzer} study of SNRs in the Galactic plane observed filamentary structure of infrared line emission in W49B from both shocked ionic and molecular gas \citep{r06}.  

\begin{figure}
\epsscale{0.5}
\plotone{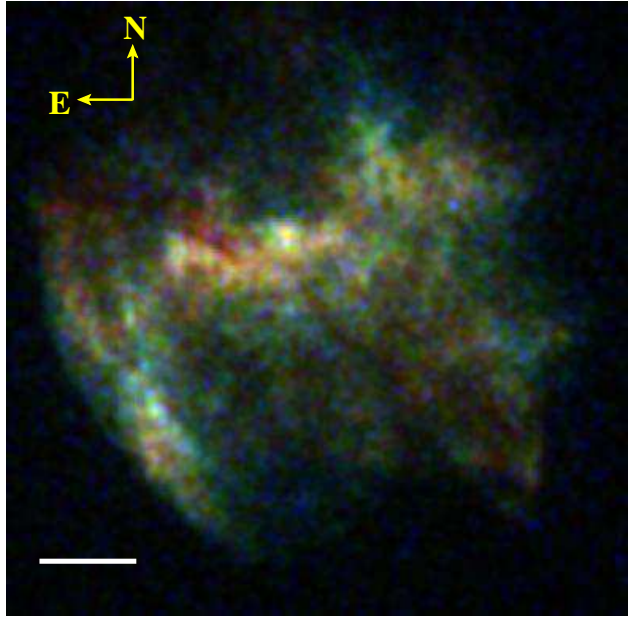}
\caption{A composite image of W49B as observed with {\it Chandra} ACIS in the soft band (0.3--1.50 keV; blue), medium band (1.5--2.5 keV; green), and hard band (2.5--8.0 keV; red). The image has been smoothed using a Gaussian kernel of width $\sigma$ = 3 pixels along each axis; the scale bar is 100 pixels long. The remnant has a bright central bar, with prominent plumes on both sides. The hard X-ray emission generally appears enclosed by the soft- and medium-band emission, except for the prominent hard X-ray emission in the Northwest region of the remnant. Soft-band emission is weak compared to the other bands because it is prefentially absorbed through the Galactic plane.}
\label{fig:three}
\end{figure}

W49B was first detected in X-rays by the {\it Einstein Observatory} \citep{p84}. An {\it EXOSAT} observation soon after revealed intense iron line emission from the source. {\it ASCA} spectra confirmed the high abundances of several metals in W49B \citep{p84, h00}, and the emission was described best by two distinct thermal components of optically-thin plasma. Given the remnant's relatively young age (\hbox{$\sim$ 1000 years:} Pye et al. 1984; Smith et al. 1985), the line emission was attributed primarily to the ejecta from the progenitor star. Recent high-resolution X-ray images (see Figure~\ref{fig:three}) and spectra from {\it Chandra} and \xmmn\ \citep{m06,k07} have shown that the metals have remarkable spatial distributions. The emission is elongated in a centrally-bright bar that is enriched in iron and nickel and has two plumes at its Eastern and Western edges that are perpendicular to the axis of this bar. The Eastern part of the remnant terminates sharply, while the Western portion of W49B is remarkably diffuse. Additionally, the Western regions have weak iron emission, whereas the lower-$Z$ elements have significant emission in these areas \citep{m06}. 

Two potential scenarios may account for the complex morphology of W49B (Miceli et al. 2006, 2008; Keohane et al. 2007). The high-metallicity and barrel-shaped morphology could be signatures of a bipolar explosion (possibly a hypernova) of a massive star. Conversely, a complex environment around a typical, symmetric supernova explosion could give rise to inhomogeneous heating to X-ray emitting temperatures. The latter is an appealing model as it is a commonly observed phenomenon (e.g., SNR G292.08+1.8: Park et al. 2004). However, this scenario would require unrealistic ejecta masses \citep{k07}. The bipolar origin of W49B is problematic as well because the predicted rate of such events is only 1 per $10^5$ years per galaxy \citep{po04}. Thus, it is uncertain presently which scenario is the correct interpretation of W49B's exotic shape. 

In order to break the degeneracy of environment versus explosion effects, quantitative methods can be applied to the existing data. The {\it Chandra} images of W49B contain a wealth of physical information that can be extracted using appropriate mathematical anaylses. Measurement of elemental segregation and mixing are important clues to explosion and evolution histories. Additionally, a rigorous description of X-ray substructure enables much more reliable estimates of elemental abundances and temperature variations than any previous work. Together, these calculations can distinguish the correct interpretation of asymmetries (anisotropic ejection or heating inhomogeneities), one of the biggest outstanding questions in SNR studies today \citep{c04}.

The anomalous spatial distribution and substructure of W49B make it an ideal source to test our methods. We aim to define quantitatively the spatial and spectral features of W49B as a means to extract physical insights about this enigmatic source. 

\section{Quantitative Methods for Imaging Analyses}

In this section, we describe the techniques that can be used to quantify X-ray morphology of extended sources. We present their mathematical formalism as well as examples of their application to synthetic data. 

\subsection{Power-Ratio Method}

The power-ratio method (PRM) was developed initially to quantify the X-ray morphology of galaxy clusters observed with {\it ROSAT} as a probe of their dynamical states (Buote \& Tsai, 1995, 1996). The technique was extended to {\it Chandra} observations to measure the evolution of cluster morphology as a function of redshift \citep{j05}. The PRM measures sensitively asymmetries in an image via calculation of the multipole moments of the X-ray surface brightness in a circular aperture. We present an overview of this technique below, and we refer the reader to the above papers for the rigorous details and examples of its application to synthetic data and X-ray observations. 

The PRM is derived similarly to the multipole expansion of the two-dimensional gravitational potential interior to radius $R$:

\begin{equation}
\Psi(R,\phi) = -2Ga_0\ln\left({1 \over R}\right) -2G
\sum^{\infty}_{m=1} {1\over m R^m}\left(a_m\cos m\phi + b_m\sin
m\phi\right). \label{eqn.multipole}
\end{equation}

\noindent
where the moments $a_m$ and $b_m$ are
\begin{eqnarray}
a_m(R) & = & \int_{R^{\prime}\le R} \Sigma(\vec x^{\prime})
\left(R^{\prime}\right)^m \cos m\phi^{\prime} d^2x^{\prime}, \nonumber \\
b_m(R) & = & \int_{R^{\prime}\le R} \Sigma(\vec x^{\prime})
\left(R^{\prime}\right)^m \sin m\phi^{\prime} d^2x^{\prime}, \nonumber
\end{eqnarray}

\noindent
$\vec x^{\prime} = (R^{\prime},\phi^{\prime})$, and $\Sigma$ is the surface mass density. For our imaging analyses, the X-ray surface brightness replaces surface mass density in the power ratio calculation. 

The powers of the multipole expansion are obtained by integrating the magnitude of $\Psi_m$ (the \textit{m}th term in the multipole expansion of the potential) over a circle of radius $R$,

\begin{equation}
P_m(R)={1 \over 2\pi}\int^{2\pi}_0\Psi_m(R, \phi)\Psi_m(R, \phi)d\phi.
\end{equation}

\noindent
Ignoring the factor of $2G$, this equation reduces to 

\begin{equation}
P_0=\left[a_0\ln\left(R\right)\right]^2 \nonumber
\end{equation}
\begin{equation}
P_m={1\over 2m^2 R^{2m}}\left( a^2_m + b^2_m\right). \nonumber
\end{equation}

The moments $a_{m}$ and $b_{m}$ (and consequently, the powers $P_{m}$) are sensitive to asymmetry in the X-ray surface brightness distribution, and the higher-order terms measure asymmetries at successively smaller scales relative to the remnant size. To normalize with respect to flux, we divide the powers by $P_{0}$ to form the power ratios, $P_m/P_0$. $P_{1}$ approaches zero when the aperture is centered on the centroid of an image, so we have set the origin in all analyses to the centroid of each individual image. In this case, morphological information is given by the higher-order terms $P_2/P_0$, $P_3/P_0$, and $P_4/P_0$. The quadrupole power $P_{2}$ is sensitive to the ellipticity of a source. The moment $P_{3}$ measures deviations from mirror symmetry (such as triangular morphology), and it is not affected by mirror-symmetrical features. The power $P_{4}$ is correlated with $P_{2}$ but is more sensitive to smaller-scale structure than $P_{2}$. Thus, the power ratios provide complementary measures of X-ray morphology in extended sources. 

\subsection{Correlation-Length Analysis}

Two-point correlation has been applied to measure galaxy clustering (e.g., Davis \& Peebles 1983) as well as to map the anisotropies of the Cosmic Microwave Background (e.g., Kogut et al. 1996). We have adapted this method to characterize spatial distribution and segregation between elements in an X-ray source, and we call our modified technique correlation-length analysis (CLA). 

CLA measures essentially whether two images have statistically similar spatial distributions. We define the correlation length of each pixel in one image as the minimum distance ($>$0 pixels) to another pixel of equal intensity in a second image. The two images are normalized to have the same maximum intensity, and pixel values are rounded to give integer counts throughout the images. We calculate the correlation length for each pixel, and we create maps of these values to identify regions with and without particular emission features. 

Additionally, we determine whether two images have similar overall spatial distributions by calculating the difference between the cumulative distributions of their correlation lengths to the cumulative distributions of each of their auto-correlations. If the two images have distinct spatial distributions, the correlation lengths between them will be disparate from those of their auto-correlations. Thus, the difference in the trends of correlation length values provides a measurement of the similarity of two images. We perform auto-correlations on mock images produced via a Monte Carlo approach to estimate the uncertainty of the calculation. 

\begin{figure}
\epsscale{0.8}
\plotone{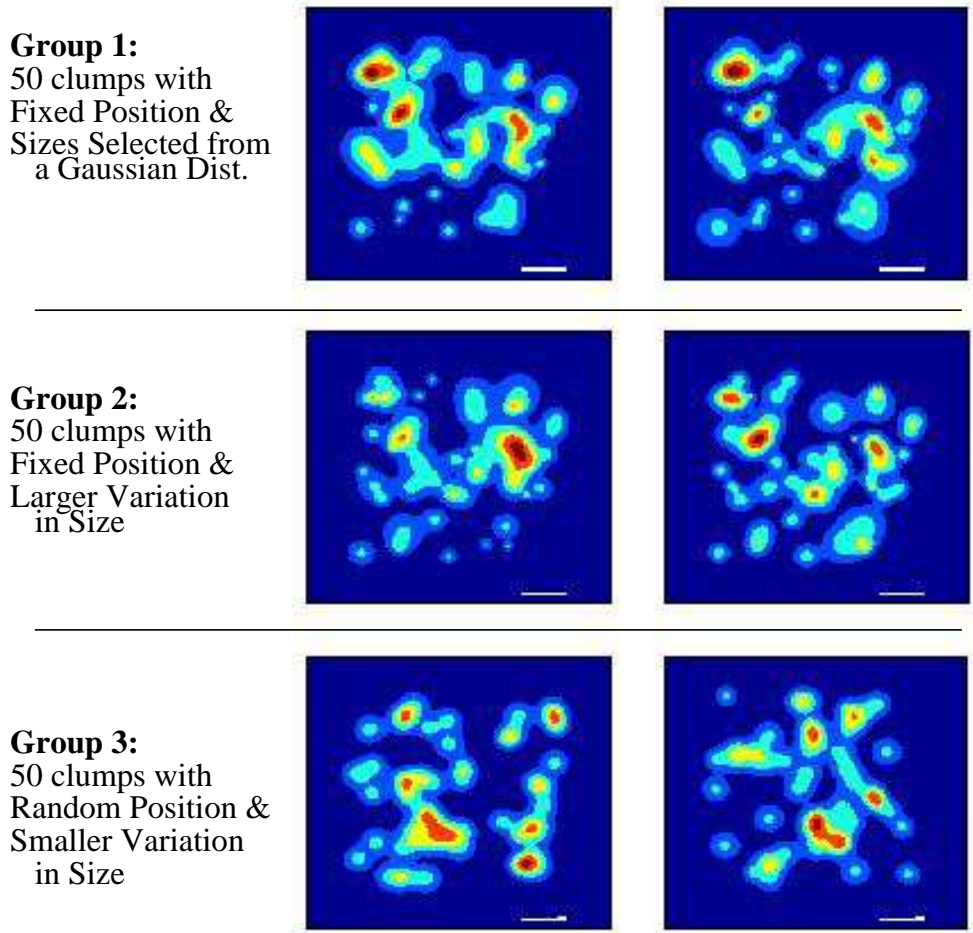}
\caption{Examples of synthetic data used in CLA. The first set of ten synthetic images (Group 1; examples in top row) contains 50 clumps of fixed position with sizes chosen from a Gaussian distribution with mean $\mu_{{\rm size}}$ = 10 pixels and a standard deviation $\sigma_{{\rm size}}$ = 3 pixels. The second set of ten images (Group 2; examples in middle row) is identical to the first group except the standard deviation in size was increased (to $\sigma_{{\rm size}}$ = 4 pixels). The third set of ten images (Group 3; examples shown in bottom row) has 50 clumps with randomly-assigned positions and sizes chosen from a Gaussian distribution with smaller standard deviation, \hbox{$\sigma_{{\rm size}}$ = 1 pixel}. The scale bar is fifty pixels in size. We performed CLA on every combination (ninety different pairings) of the ten images in each group and produced the three mean correlation-length CDFs and their 1-$\sigma$ ranges shown in Figure~\ref{fig:cl}.} 
\label{fig:examplecl}
\end{figure}

\begin{figure}
\epsscale{0.8}
\plotone{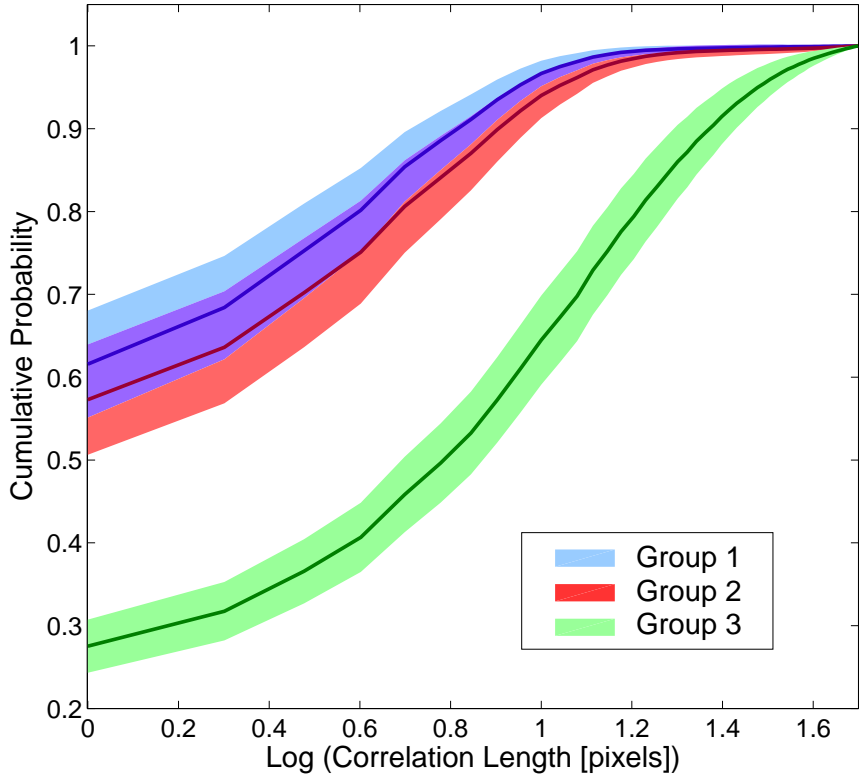}
\caption{Correlation-length cumulative-distribution functions (CDFs) and their 1-$\sigma$ confidence ranges for the three groups of synthetic data (Group 1: blue; Group 2: red; Group 3: green). Since the CDFs of Groups 1 and 2 overlap, the two have statistically similar spatial distributions, indicating that CLA is not sensitive to moderate changes in clump size. By constrast, Group 3 has a very distinct CDF curve since its clumps have randomly-selected locations (unlike Groups 1 and 2). These results suggest that CLA is a good means to determine if images have similar distributions, regardless of whether the clumps are the same size.} 
\label{fig:cl}
\end{figure}

We test the utility of CLA by applying the method to three groups of ten synthetic images (Figure~\ref{fig:examplecl} has examples and a mathematical description of the datasets). Groups 1 and 2 have similar distributions and represent auto-correlations, while Group 3 reflects comparison of disjoint distributions. We performed CLA on every combination (ninety different pairings) of the ten images within each group and produced the three mean correlation-length CDFs and their 1-$\sigma$ ranges shown in Figure~\ref{fig:cl}. 

To determine whether the size of substructure influences CLA results, Groups 1 and 2 have clumps at fixed location but with scales selected from Gaussian distributions of different variance: the clump-size distribution of Group 2 has $56\%$ larger variance than that of Group 1. Comparing their resulting CDFs, we find that the confidence ranges overlap for these datasets, with an average separation of 0.5$\sigma$. Thus, images in Groups 1 and 2 have statistically similar spatial distributions, indicating that clump-size variation does not hinder CLA from distinguishing homogeneous populations. 

To ascertain how clump position affects CLA results, clump locations of Group 3 datasets were selected randomly while the clump sizes did not vary much (with 11\% of the Group 1 variance). In this case, many pixels have large correlation lengths, so the Group 3 CDF is very disparate from those of Groups 1 and 2 with average separations of 3.1$\sigma$ and 2.7$\sigma$, respectively. Thus, Group 3 has a statistically distinct spatial distribution from Groups 1 and 2, and we conclude that CLA is a good measure of whether substructures have similar locations. Therefore, CLA is well-suited to compare morphologies within and between extended sources. 

\subsection{Wavelet-Transform Analysis}

Wavelet-transform analysis (WTA) was applied successfully to {\it ROSAT} and {\it Einstein} data to extract small-scale X-ray structure of galaxy clusters \citep{g95}. With the superior spatial resolution of {\it Chandra}, this technique can be extended to characterize precisely the size and distribution of emitting regions in X-ray sources. A wavelet basis enables scale as well as spatial localization, thus WTA can identify substructure, its extent, and its position. 

\subsubsection{Mathematical Formalism}

A wavelet-transformed image is a decomposed image of a signal's intensity (from herein, power) measured at the scale of a filter size. Mathematically, a wavelet transform $w$ is the correlation of a signal $s(x,y)$ in an image with the analyzing wavelet function $g(x,y)$:

\begin{equation}
w(x,y,a) = s (x,y) \otimes \frac{1}{a} g \bigg(\frac{x}{a},\frac{y}{a} \bigg),
\end{equation}

\noindent
where $a$ is the scale (or width) of the wavelet transform. For astronomical images, a radial Mexican-hat function (the normalized second derivative of a Gaussian function) is the optimal wavelet to use because of its similar shape to Gaussian signals and it removes flat features (like diffuse emission) since it has a zero mean \citep{s90}. In the analysis that follows, we use a radial Mexican-hat wavelet $g(\frac{x}{a},\frac{y}{a})$ of the form

\begin{equation}
g(\frac{x}{a},\frac{y}{a}) = \bigg(2 - \frac{x^{2}+y^{2}}{a^{2}} \bigg) e^{-(x^{2}+y^{2})/2a^{2}}.
\end{equation}

Wavelet-transformed images are produced by calculating $w(x,y,a)$ for each pixel $(m,n)$ in a raw image:

\begin{equation}
w(m,n,a) = \frac{1}{a} \sum c_{ij} g \bigg( \frac{x_i - x_m}{a} , \frac{y_j - y_n}{a} \bigg),
\end{equation}

\noindent
where $c_{ij}$ is the number of counts in the ($i,j$) pixel. If we assume the number of counts in a given pixel is independent of the counts in nearby pixels, we can use Poisson statistics to calculate the variance $q(m, n, a)$ of $w(m, n, a)$:

\begin{equation}
q(m,n,a) = \frac{1}{a^2} \sum d_{ij} g^{2} \bigg( \frac{x_{i} - x_{m}}{a} , \frac{y_{i} - y_{n}}{a} \bigg), 
\end{equation}

\noindent
where $d_{ij}$ is the variance of $c_{ij}$ in the $(i, j)$ pixel of a raw image. 

Essentially, $w$ measures the summed intensity enclosed by the area of the Mexican hat. Thus, the size of an individual source can be characterized by the scale where the convolution of the wavelet and a signal reaches a maximum. To exemplify this point mathematically, we assume that a source has an isotropic Gaussian shape with signal $s(x,y,\sigma)$ given by

\begin{equation}
s(x,y,\sigma) = \frac{I}{2 \pi \sigma^2} \exp \bigg[ - \frac{(x - x_o)^2 + (y - y_o)^2}{2 \sigma^2} \bigg],
\end{equation}

\noindent
where $I$ is the intensity of the source located at $(x_o,y_o)$. Using Eq. 5 and Eq. 6, the wavelet coefficient $w(x_o, y_o, a)$ at this position is

\begin{equation}
w(x_o, y_o, a) = \frac{2 I}{a} \bigg( 1 + \frac{\sigma^2}{a^2} \bigg)^{-2}. 
\end{equation}

\noindent
If we divide Eq. 10 by $a$, we find that

\begin{equation}
\frac{w}{a} = \frac{2 I}{a^2} \bigg( 1 + \frac{\sigma^2}{a^2} \bigg)^{-2}. 
\end{equation}

\noindent
Eq. 11 has a global maximum at $a_{{\rm max}}$ = $\sigma$. Consequently, by identifying the peak $w/a$ value at the central pixel of an emitting region, we measure its characteristic size. 

\subsubsection{Application to Synthetic Data}

\begin{figure}
\epsscale{0.9}
\plotone{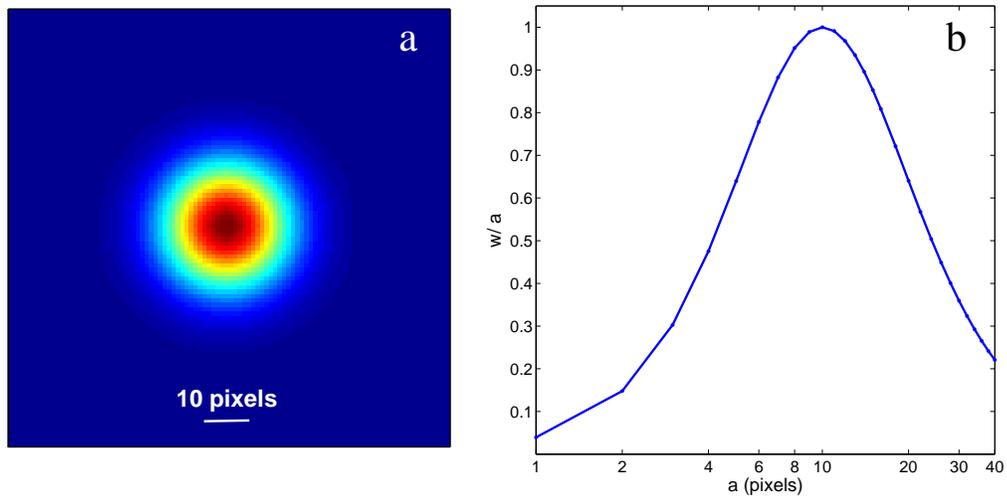}
\caption{An example of wavelet-transform analysis. a: Synthetic data of a single two-dimensional Gaussian function with a width $\sigma$ = 10 pixels. b: Plot of the relative power ($w/a$) as a function of Mexican-hat radius $a$ for the central pixel of the clump. Increased values of $w/a$ indicating that more power is emitted at a given scales. Thus, local and absolute maxima denote the size of an emitting region. Indeed, the peak of the plot occurs at $a_{{\rm max}}$ = $\sigma$.}
\label{fig:gauss}
\end{figure}

To demonstrate the utility of WTA for astronomical images, we apply the method to relevant synthetic datasets. We begin by using WTA to extract the size of a single two-dimensional Gaussian function (from herein, a clump) with a width $\sigma$. Figure~\ref{fig:gauss}a shows this synthetic data, and Figure~\ref{fig:gauss}b gives the plot of $w/a$ versus $a$ (intensity per unit area as a function of scale; from herein, WTA plot) for the central pixel of the clump. Larger $w/a$ values correspond to more power emitted at the given scales, so local and absolute maxima indicate the size of emitting regions. Indeed, the peak $a_{{\rm max}}$ of the WTA plot occurs at the exact value of $\sigma$. Therefore, the method can measure accurately the size of an individual emitting region or substructure.

\begin{figure}
\epsscale{1.0}
\plotone{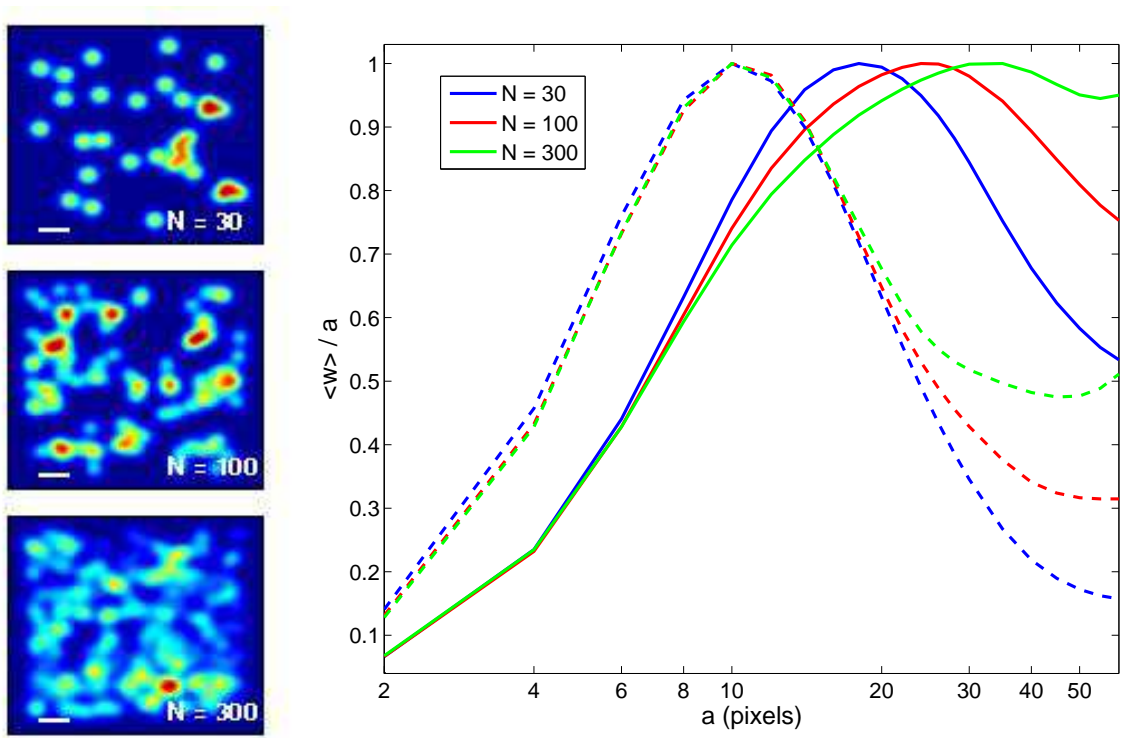}
\caption{Synthetic datasets (left) with different numbers $N$ of randomly distributed Gaussians of equal width $\sigma$ = 10 pixels: $N$ = 30 clumps, 100 clumps, and 300 clumps. All three arrays are the same size, 300 $\times$ 300 pixels, and the scale bar is fifty pixels in size. An increase in $N$ corresponds to emission over more surface area (increased filling factor), so the power at larger scales will rise as a function of $N$. The averaged WTA plot ( $\langle w \rangle /a$ versus $a$; right, solid lines) reflects this relationship: the $N$ = 30, 100, and 300 images had $a_{{\rm max}}$ of 16 pixels, 22 pixels, and 26 pixels, respectively. Additionally, the maxima become less prominent with larger $N$ because of this increased filling factor. Thus, the peak of the aveaged WTA plot is sensitive to the filling factor of an emitting region. The averaged WTA plot for isolated clumps (right, dashed lines) accurately reflects the scales of those structures. The fraction of identified isolated clumps decreases with larger filling factors: 24 isolated clumps were identified in the $N$ = 30 clumps case, whereas 43 isolated clumps were identified in the $N$ = 300 case.}
\label{fig:nsigma}
\end{figure}

By applying WTA to all the pixels in an image and summing their results for each scale, the method can be utilized to calculate the intensity profile of an entire extended source. As an example, we have applied WTA to all pixels in three synthetic datasets with randomly-distributed clumps of equal width (Figure~\ref{fig:nsigma}, left). The number of clumps in these images are varied to demonstrate the effects of increased emitting surface area, or filling factor. Figure~\ref{fig:nsigma} (right; solid lines) displays the averaged WTA plot ($\langle w \rangle /a$ versus $a$) for the three datasets. The curves have similar shape as Figure~\ref{fig:gauss}, and increased $\langle w \rangle /a$ values correspond to more power emitted at a particular scale over an entire image. As the number of clumps increases and they overlap, larger agglomerates are formed. This effect increases the surface area of contiguous emitting regions, and the amount of power measured at large scales to rise. Thus, increased numbers of clumps shift the peak $a_{{\rm max}}$ to higher values, and the turnover in $\langle w \rangle /a$  becomes less prominent. 

In images with many clumps, we can measure the scale of isolated structures using the WTA plot for their central pixels. However, larger filling factors cause individual clumps to agglomerate, and the number of segregated structures decrease as a result. In Figure~\ref{fig:nsigma} (right; dashed lines), we plot the averaged WTA plot for the individually identified clumps in the three synthetic datasets of randomly-distributed clumps of equal width (Figure~\ref{fig:nsigma}, left). We define the centroid of individual clumps as the locations with local maxima in $w/a$ separated at least three pixels from another local maximum. In all cases, the algorithm successfully reproduces the scale of the isolated structures. Generally, our ability to identify individual clumps decreases with greater emitting surface area (see Figure~\ref{fig:nsigma}). Consequently, WTA probes an extended source's filling factor as well as the size of its individual emitting regions. 

\begin{figure}
\epsscale{1.0}
\plotone{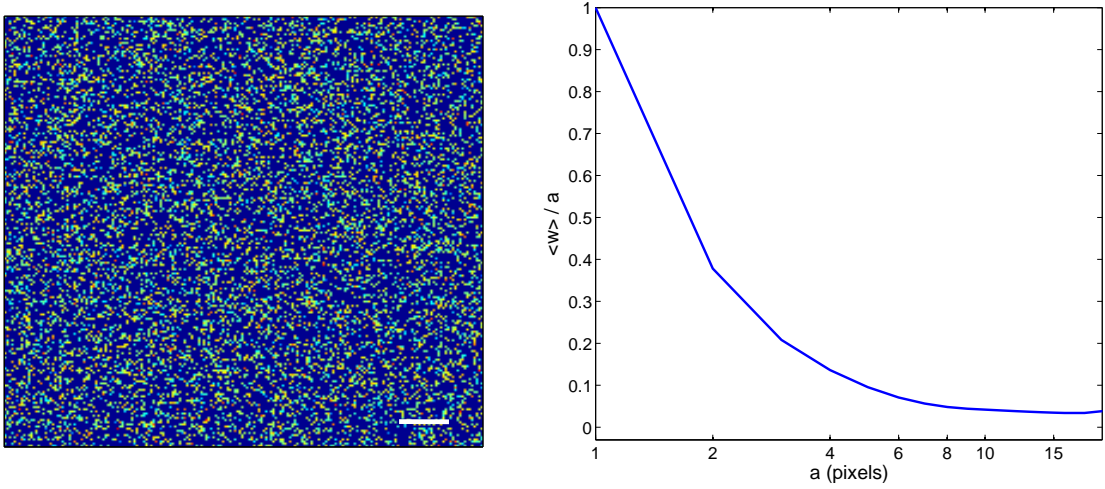}
\caption{Synthetic data of pure noise (left) and the resulting averaged WTA plot (right). The scale bar on the synthetic data is fifty pixels in size. The intensity of each pixel with noise was chosen from a Poisson distribution with standard deviation $\sigma_{{\rm noise}}$ = 3 counts. The power profile peaks at $a_{{\rm max}}$ = 1 pixel, indicating that noise contributes the most power at the single pixel scale.}
\label{fig:purenoise}
\end{figure}

\begin{figure}
\epsscale{1.0}
\plotone{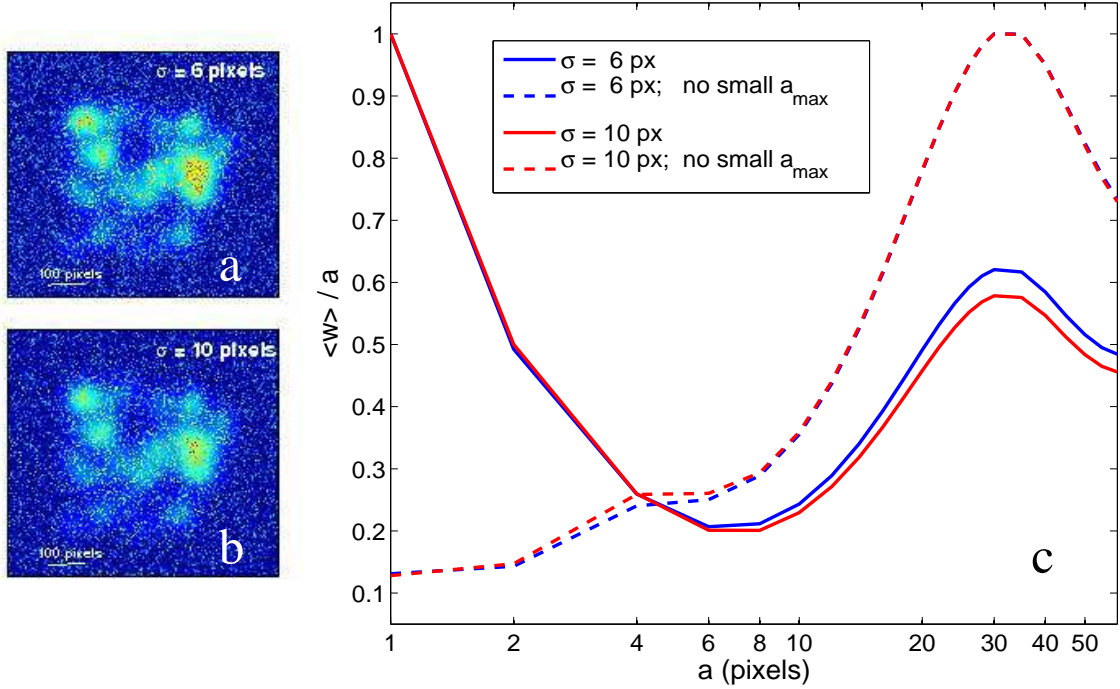}
\caption{Examples of synthetic data of pure noise and 50 clumps combined (a and b) and their WTA results (c). The clump sizes were chosen from Gaussian distributions with a mean $\mu_{{\rm sizes}}$ = 30 pixels and standard deviation $\sigma_{{\rm sizes}}$ = 6 pixels (a) or $\sigma_{{\rm sizes}}$ = 10 pixels (b). The averaged WTA plot (c) for the datasets overall (shown with solid lines) have global maxima at \hbox{$a_{{\rm max}}$ = 1 pixel} because of the noise. However, if we ignore pixels with global maxima at $a_{{\rm max}} = 1$ or 2 pixels, the contribution from noise is drastically reduced and the global maxima become \hbox{$a_{{\rm max}} \mu_{{\rm sizes}}$ = 30 pixels}. The method extracted successfully the characteristic size of the clumps, regardless of their variance in scale.} 
\label{fig:npd}
\end{figure}

Real astronomical images have noise that will naturally influence these WTA results. To investigate the effects of noise, first we apply our techniques to synthetic data of pure noise (Figure~\ref{fig:purenoise}, left). The noise intensity in each pixel was selected from a Poisson distribution. Figure~\ref{fig:purenoise} (right) shows the resulting averaged WTA plot; the absolute maximum of the WTA curve is at $a$ = 1 pixel. Since the noise is distributed on a pixel-by-pixel basis, it is natural that the most power is emitted at the single-pixel scale. We then add the pure noise (Figure~\ref{fig:purenoise}, left) to synthetic data of randomly-distributed clumps to simulate realistic images of an extended astronomical source. Clump sizes were selected from Gaussian distributions, and images were produced using different variances in the clump-size distribution to investigate the effects of substructure on WTA results. We calculate the averaged WTA plot for the datasets; two example images and their results are shown in Figure~\ref{fig:npd}. All curves have global maxima at $a_{{\rm max}} = 1$ pixel since the noise is dominant. However, we can reduce its contribution by ignoring all pixels with global maxima at the smallest scales. Using this approach, the global maxima shift toward the scale of mean clump size since the most power resides there when noise is excluded. We determine that the variance in clump size did not alter the WTA results either: the intensity profile is similar regardless of how divergent individual clumps are from the average scale. 

\begin{figure}
\epsscale{1.0}
\plotone{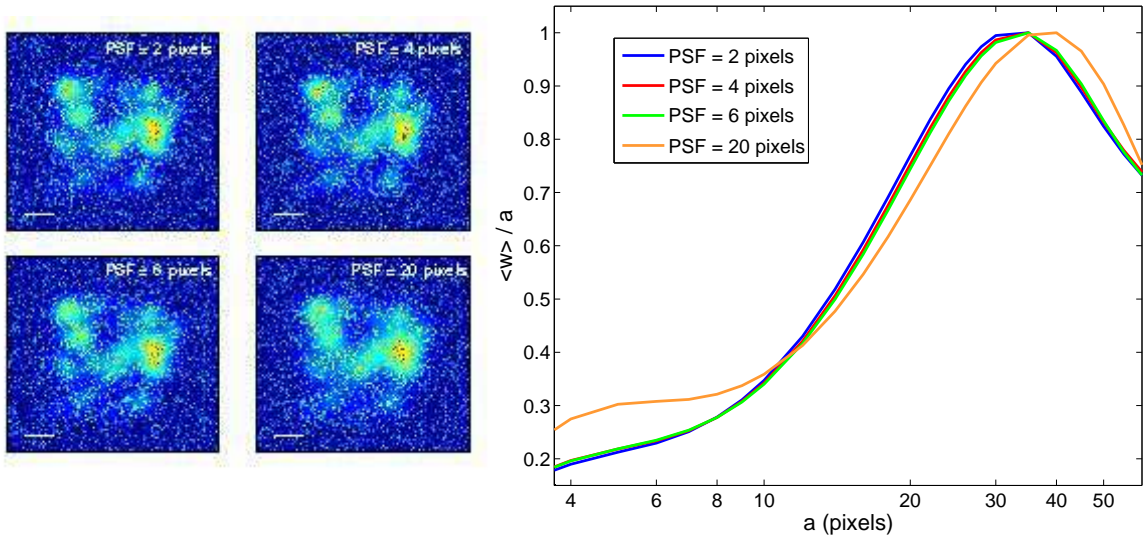}
\caption{Synthetic data of noise and clumps combined (Figure 9a), convolved with Gaussian functions of widths $\sigma$ = 2, 4, 6, and 20 pixels to simulate PSF effects on the clumps in an astronomical image. Generally, the scale of the convolving Gaussians does not shift the WTA peak, and larger Gaussians decrease the power contribution of noise at the smallest scales.} 
\label{fig:convolve}
\end{figure}

Instrumental effects may alter WTA results: in particular, an image's resolution is limited by the point-spread function (PSF) of the detector. To investigate how an instrument's PSF changes the scale where most power is emitted, we convolved the data in Figure~\ref{fig:npd}a with Gaussian functions and then add noise. The resulting four images are shown in Figure~\ref{fig:convolve}, along with their averaged WTA plot. Generally, the convolution did not influence the intensity profile measurement. We find that the results are unaffected for all PSFs 60\% or less of their scale, and the maxima in the WTA plot increase with PSFs larger than this fraction. 

Based on the results from the synthetic data presented above, we conclude that WTA is a useful technique to extract the individual sizes of emitting regions and to measure a source's filling factor, an important geometric parameter. Noise increases the power extracted at the smallest scales, but we can limit this contribution by ignoring pixels with global maxima of a single pixel. Generally, large variation in clump sizes does not reduce the accuracy in determining their intensity profiles, and WTA measures the various emitting regions' scales necessary to produce a given surface brightness. Thus, WTA is a useful tool to probe the substructure of extended sources, such as supernova remnants. 

\subsection{Uncertainties and Limitations of the Methods}

Although the methods are successful on these synthetic data, the techniques have uncertainties and limitations that should be considered. The power-ratio method necessitates an accurate determination of an image's centroid; this calculation requires a source to have a minimum flux and minimum signal-to-noise ratio (S/N). As shown by \cite{j05}, circular apertures with greater than 500 net counts and S/N greater than 3.0 are sufficient to locate the centroid for unbinned {\it Chandra} data. Furthermore, to obtain accurate net moments of extended sources, a few criteria should be satisfied: point sources must be removed, the background moment must be much less than the source moment, and the source must fit entirely on the detector. \cite{j05} also notes that low S/N images with highly symmetric/circular sources will have systematically higher power ratios due to noise. A Monte Carlo approach can be employed to estimate the uncertainty associated with noise (as described in $\S$3.2). 

In the case of the CLA, the primary limitation is associated with the normalization of the two images to have maximum equal intensity. If the images have very disparate S/N, the normalization procedure will alter the relative quality of the images. Namely, if the maximum intensity of the high S/N image is scaled down to match that of the low S/N image, weaker features in the high S/N image would be removed artificially and thus ignored in the CLA results. Therefore, the method is best applied to images with similar intensity prior to normalization. As with the PRM, a Monte Carlo approach can be used to measure the the effects of noise on the resulting cumulative-distribution function (as described in $\S$3.3). 

Unlike the other methods, noise is not a significant source of uncertainty in WTA provided that certain criteria are met. We have shown in $\S$2.3.2 that we can limit the affects of noise by ignoring pixels with global maxima of a single pixel. Generally, we find that with this procedure, the averaged WTA plot is roughly independent of S/N for S/N $>$ 2.0 as long as $a_{\rm max}$ is much greater than 1 pixel ($a_{\rm max}$ $>$ 8 pixels). If emitting regions have S/N and $a_{\rm max}$ below these values, a Monte Carlo approach can be employed to estimate the uncertainty introduced by noise. 

For WTA, the principal limitation arises from increased filling factor. In $\S$2.3.2, we demonstrated that our ability to identify individual clumps diminishes with greater filling factor (see Figure~\ref{fig:nsigma}). The reason for this difficulty is that when clumps agglomerate into larger structures, it becomes challenging to distinguish the central pixel of that emitting region. Thus, isolated clumps in sources with low filling factors are the best candidates for WTA to measure indiviudual clump sizes. Large filling factors negatively influence the averaged WTA results as well: the turnover in the averaged WTA plot becomes less prominent with increased emitting surface area in an extended source. Generally, we find that images with filling factors above $\sim$30\% do not produce a statistically significant $a_{\rm max}$. 

\section{Data Analysis}

In the subsequent sections, we demonstrate the application of the three techniques of $\S$2 to a particular SNR, the galactic remnant W49B. 

\subsection{Observations and Data Preparation}

We analyzed a 55-ks archival {\it Chandra} ACIS observation of W49B (ObsID = 117). The remnant was observed with the backside-illuminated S3 chip in the Timed-Exposure Faint Mode. Data reduction and analysis was performed using the {\it Chandra} Interactive Analysis of Observations ({\sc ciao}) Version 3.4. We followed the {\sc ciao} data preparation thread to reprocess the Level 2 X-ray data: we adjusted the charge transfer inefficiency and time-dependent gain, and we randomized the PHA distribution and the event positions. {\it ASCA} grade 0, 2, 3, 4, and 6 events were used in the analysis, and good time intervals were applied. Data below \hbox{0.3 keV} and above \hbox{8 keV} were ignored. Throughout the paper, we assume the distance $D$ to W49B is $D$ = 8 kpc; this value implies an ACIS pixel size of 0.492\arcsec\ $\approx$ 0.019 pc. 

\begin{figure}
\epsscale{0.8}
\plotone{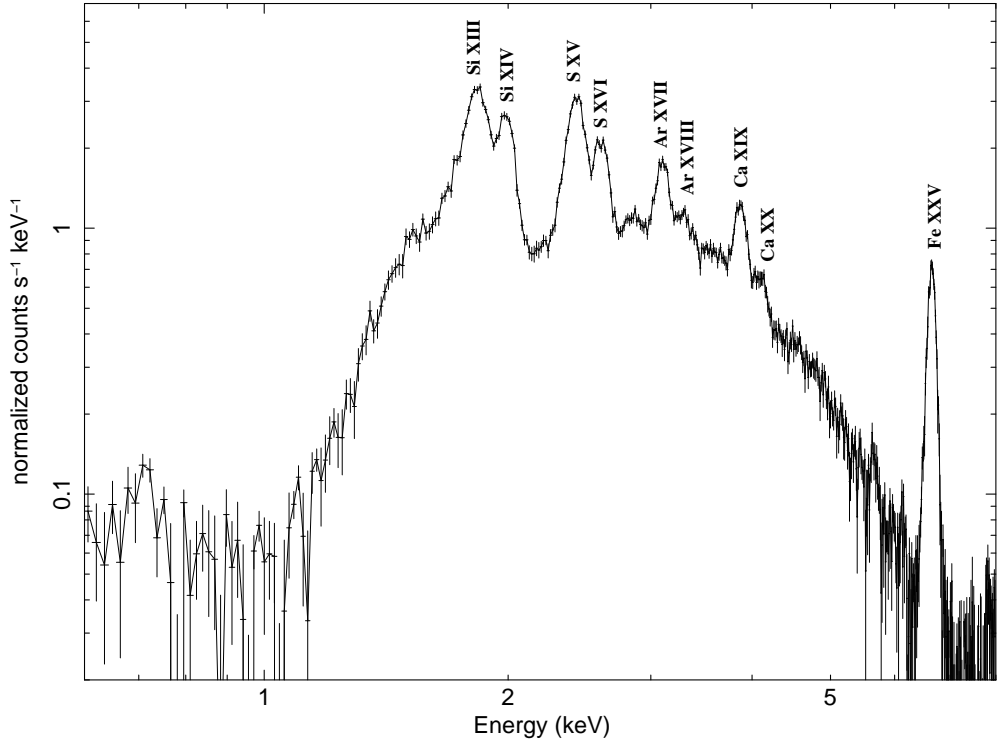}
\caption{Background-subtracted global X-ray spectrum of W49B obtained with {\it Chandra} ACIS. The spectrum is composed of a thermal bremsstrahlung continuum and many prominent emission lines, including blends of Si, S, Ar, Ca, and Fe. No emission features are detected below $\approx$1.7 keV because the source is heavily absorbed since it is viewed through the Galactic plane.}
\label{fig:spectrum}
\end{figure}

We extracted the global {\it Chandra} ACIS X-ray spectrum of W49B (shown in Figure~\ref{fig:spectrum}) using the {\sc ciao} command {\it specextract}, and $\approx$2.4$\times$10$^{5}$ total counts were detected from the remnant. The X-ray spectrum is quite complex: it has a thermal bremsstrahlung continuum and numerous emission features. Prominent blends include both He-like ions and the H-like ions of silicon, sulfur, argon, and calcium. Additionally, a strong iron line is detected. Modeling of \xmmn\ spectra from W49B suggests at least two plasma components are necessary to adequately fit these diverse emission features since they are produced at different electron temperatures \citep{m06}. 

\begin{figure}
\epsscale{0.9}
\plotone{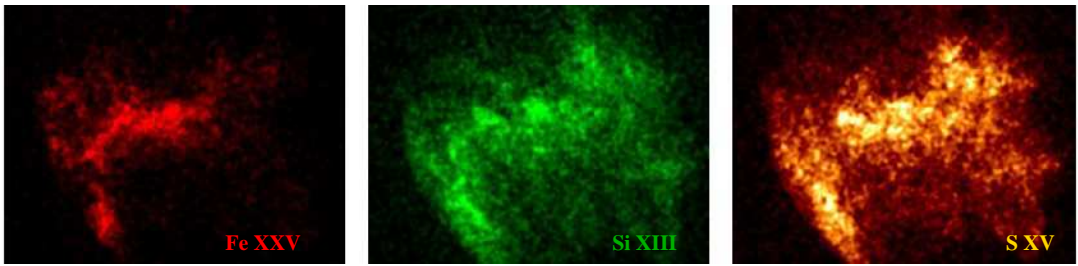}
\caption{ACIS X-ray images of Fe {\sc XXV} (left), Si {\sc XIII} (middle), and S {\sc XV} (right). Images are $\approx$250\arcsec\ across and were smoothed using a Gaussian kernel of width $\sigma$ = 3 pixels along each axis. Iron has a very distinct spatial distribution compared to the lighter ions, and all have many X-ray substructures.}
\label{fig:ions}
\end{figure}

Exposure-corrected images of the strong emission features were produced by filtering data to the narrow energy ranges of the prominent emission lines (as labeled in Figure~\ref{fig:spectrum}): Si {\sc xiii} (1.74--1.93 keV), Si {\sc xiv} (1.94--2.05 keV), S {\sc xv} (2.25--2.50 keV), S {\sc xvi} (2.52--2.71 keV), Ar {\sc xvii} (2.98--3.21 keV), Ca {\sc xix} (3.75--4.0 keV), and Fe {\sc xxv} (6.4--6.9 keV). We multiplied all the exposure-corrected images by the maxima of their exposure maps so that every image had units of counts. The spatial distribution of the elements that produce these lines is inhomogeneous (Figure~\ref{fig:ions} gives examples). The iron is much more prominent in the Eastern portion of the remnant, while silicon and sulfur are comparatively more symmetric. Additionally, all three ions have discernible X-ray substructure, particularly in the central and Western regions where clumping is evident. 

The line emission in these images will have contamination from the underlying thermal continuum, and it is necessary to remove this component prior to imaging analyses. We estimate the flux of the bremsstrahlung and lines by phenomenologically modeling the X-ray spectra from different regions of the remnant. We find that the relative flux of the thermal emission in the narrow-band images varies by only $\sim$5\% over the entire source. For example, bremsstrahlung in the energy range of the Si {\sc xiii} line contributes 35--40\% of the flux for all regions. Thus, we continuum-subtract the elemental images by reducing the counts in each pixel by the average fraction of flux from the thermal component in that spectral band. 

\subsection{Ion Symmetry}

We calculate the power ratios $P_2/P_0$, $P_3/P_0$, and $P_4/P_0$ of the continuum-subtracted, exposure-corrected images of the five strongest emission features in the W49B spectrum (Si {\sc xiii}, S {\sc xv}, Ar {\sc xvii}, Ca {\sc xix}, and Fe {\sc xxv}) using the method outlined in $\S$2.1.  These five images have $\sim$7500--25000 counts each; thus, they satisfy the criteria outlined in $\S$2.4 for accurate net moment determinations. The results are listed in Table 1. For each ion image, we determined its centroid and calculated the power ratios using a aperture radius that encloses the entire remnant, $R_{ap}$ = 150\arcsec. To account for the X-ray background, we measured the counts per pixel in a circular annulus 155\arcsec\ -- 190\arcsec\ from the centroid of each ion image. Assuming a constant background, we produced background images for each ion with the same dimensions as the source images by setting every pixel to the extracted count-per-pixel values. We calculated the moments of the source and of its background separately, and we subtracted the background moments from the source moments to obtain the powers using these net moments. 

To estimate the uncertainty in the power ratios, we follow the Monte Carlo approach employed in the X-ray cluster studies \citep{b96,j05}. Exposure-corrected images (normalized to have units of counts) are adaptively-binned using the program {\it AdaptiveBin} \citep{s01} such that all zero pixels are removed (corresponding to a minimum of $\approx$2 \hbox{counts bin$^{-1}$}) to smooth out noise. Then, noise is added back in by taking each pixel intensity as the mean of a Poisson distribution and selecting randomly a new intensity from that distribution. This process was repeated 100 times for each of the five elements, creating 100 mock images per ion. The 90\% confidence limits listed in Table 1 were obtained using the fifth highest and fifth lowest power ratios from the 100 mock images of each element. 

\begin{figure}
\epsscale{0.35}
\plotone{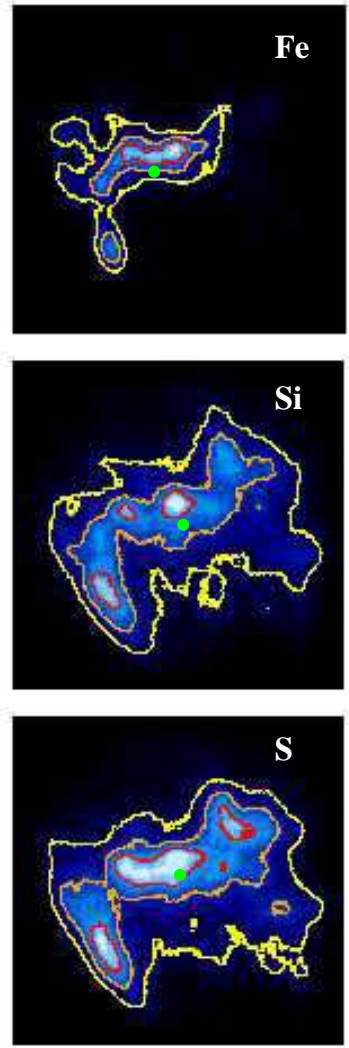}
\caption{Surface brightness contours overlayed on smoothed images of Fe{\sc xxv} (top), Si {\sc xiii} (middle), and S {\sc xv} (bottom). Contours mark 25\% (yellow), 50\% (orange), and 75\% (red) of the maximum surface brightness in each image. The green circles mark the centroids used in the power-ratio analysis; the centroid of iron is offset from the other elements by $\sim$25\arcsec. The surface brightness contours reflect the lack of iron in the Western portion of the remnant, while the other elements are comparatively more symmetrically and diffusely distributed. The power ratio method quantifies these morphologies and statistically confirms the asymmetry of iron relative to silicon, sulfur, argon, and calcium.} 
\label{fig:contours}
\end{figure}

We find that the power ratios of Fe {\sc xxv} are statistically different than the other elements. The ratio $P_2/P_0$ (a measure of ellipticity) of iron is roughly half the values obtained for Si {\sc xiii}, S {\sc xv}, Ar {\sc xvii}, and Ca {\sc xix}. Similarly, $P_4/P_0$ (a measure of ellipticity on smaller scales than $P_2/P_0$) of Fe {\sc xxv} is 2--4 times smaller than the lighter ions. However, $P_3/P_0$ (a measure of deviations from mirror symmetry) of Fe {\sc xxv} is a factor of 4--5 higher than the values of the other ions. The surface brightness contours of these ions demonstrate the reason for these results (Figure~\ref{fig:contours}). The large $P_2/P_0$ and $P_4/P_0$ of the lighter ions reflect their comparatively elongated distribution relative to Fe {\sc xxv}. The large $P_3/P_0$ of Fe {\sc xxv} arises because of its absence from the Western part of the remnant while it has bright substructure in the central and Eastern regions of W49B. The other ions have much smaller $P_3/P_0$ values since their distributions are comparatively more balanced in the Eastern and Western portions of the remnant. Therefore, the morphology of Fe {\sc xxv} is quantitatively less elongated and more asymmetric than the lighter ions. 

\subsection{Ion Distribution}

\begin{figure}
\epsscale{1.0}
\plotone{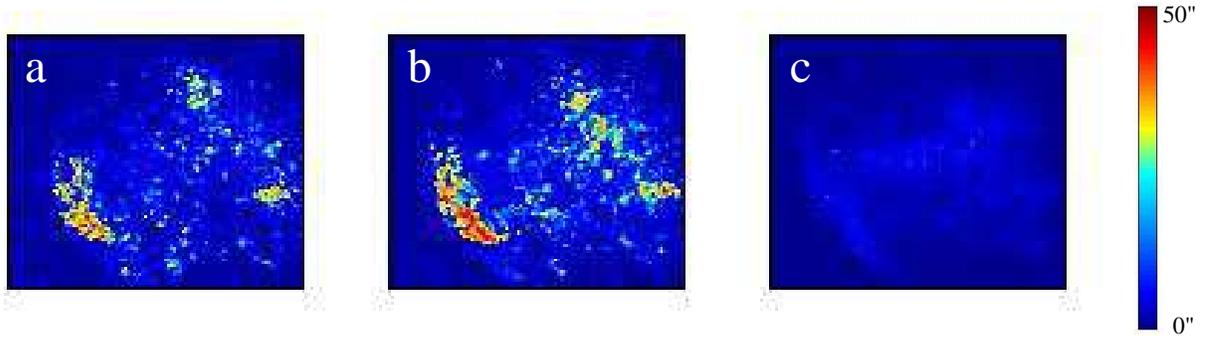}
\caption{Example maps of CLA results: silicon versus iron (a); sulfur versus iron (b); silicon versus sulfur (c). Maps are 300\arcsec\ across. Yellow and red signify large correlation lengths, i.e. regions in which the two given ions do not have statistically similar spatial distributions. Iron appears to be very segregated from silicon and sulfur, while the latter seem to be well-mixed, a result that is confirmed quantitatively in Figure~\ref{fig:cdf}.} 
\label{fig:ccmaps}
\end{figure}

We perform the correlation-length analysis (CLA) outlined in $\S$2.2 on images of the five strongest emission lines in the {\it Chandra} ACIS spectrum. All the ion images have roughly the same maximum pixel intensity ($\sim$5 counts) prior to normalization, so the limitations of CLA discussed in $\S$2.4 do not apply to the W49B data. We analyzed each combination of the five images and produced their correlation-length maps, with examples shown of the three most prominent lines (Fe {\sc xxv}, Si {\sc xiii}, and S {\sc xv}) in Figure~\ref{fig:ccmaps}. We find that all ions are well-mixed throughout the remnant, except for Fe {\sc xxv} which is largely absent from the plumes of the remnant relative to the other ions. 

\begin{figure}
\epsscale{0.8}
\plotone{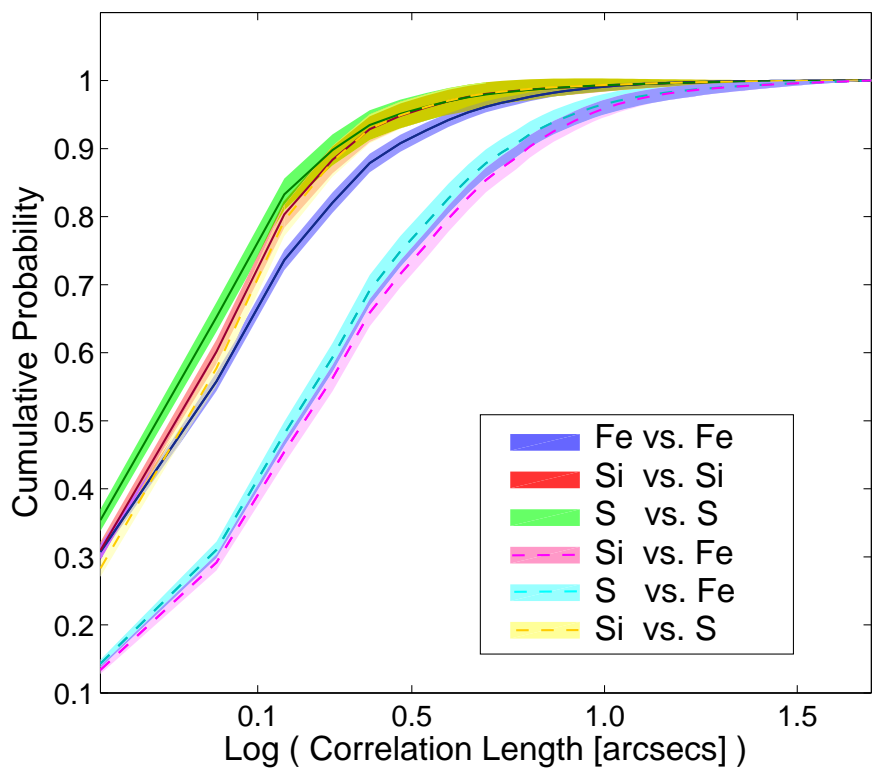}
\caption{CLA results for iron, silicon, and sulfur. Solid lines and their respective shaded regions denote the cumulative distribution functions and 1-$\sigma$ confidence ranges for iron auto-correlation (blue), silicon auto-correlation (red), and sulfur auto-correlation (green). Dashed lines and their respective shaded regions indicate the CDFs and 1-$\sigma$ ranges for silicon vs. iron (pink), sulfur vs. iron (light blue), and silicon vs. sulfur (yellow). The silicon versus sulfur case is similar to those of the ions against themselves, indicating the two elements have statistically similar spatial distributions. By contrast, the CDFs of those elements against iron are very disparate from the other curves, demonstrating quantitatively the segregation of iron relative to the other ions.} 
\label{fig:cdf}
\end{figure}

Additionally, we produced the correlation-length cumulative-distribution functions shown in Figure~\ref{fig:cdf}. 1-$\sigma$ confidence ranges of these CDFs were calculated by performing CLA on every pair of ten mock images (the same mock images as in $\S$3.2) of each ion. To determine whether the ions have similar spatial distributions, we compare the CDFs to those produced using auto-correlation (Figure~\ref{fig:cdf}). We find that the silicon-to-sulfur CDF is statistically similar to the silicon auto-correlation CDF (with an average difference of 0.007$\sigma$) and to the sulfur auto-correlation CDF (with an average difference of 0.060$\sigma$). This result indicates that Si {\sc xiii} and S {\sc xv} have the same spatial distribution and are well mixed. Conversely, Fe {\sc xxv} has a very different spatial distribution than Si {\sc xiii} and S {\sc xv}: the CDFs are very disparate from those of the auto-correlations, with average differences of 4.6$\sigma$ from Si {\sc xiii} and 3.7$\sigma$ from S {\sc xv} for correlation lengths under 10\arcsec. Thus, the iron is more segregated and disjoint from the other ions, whereas all except Fe {\sc xxv} seem to be well-mixed and similarly distributed throughout the remnant.

\subsection{X-ray Substructure}

\begin{figure}
\epsscale{1.0}
\plotone{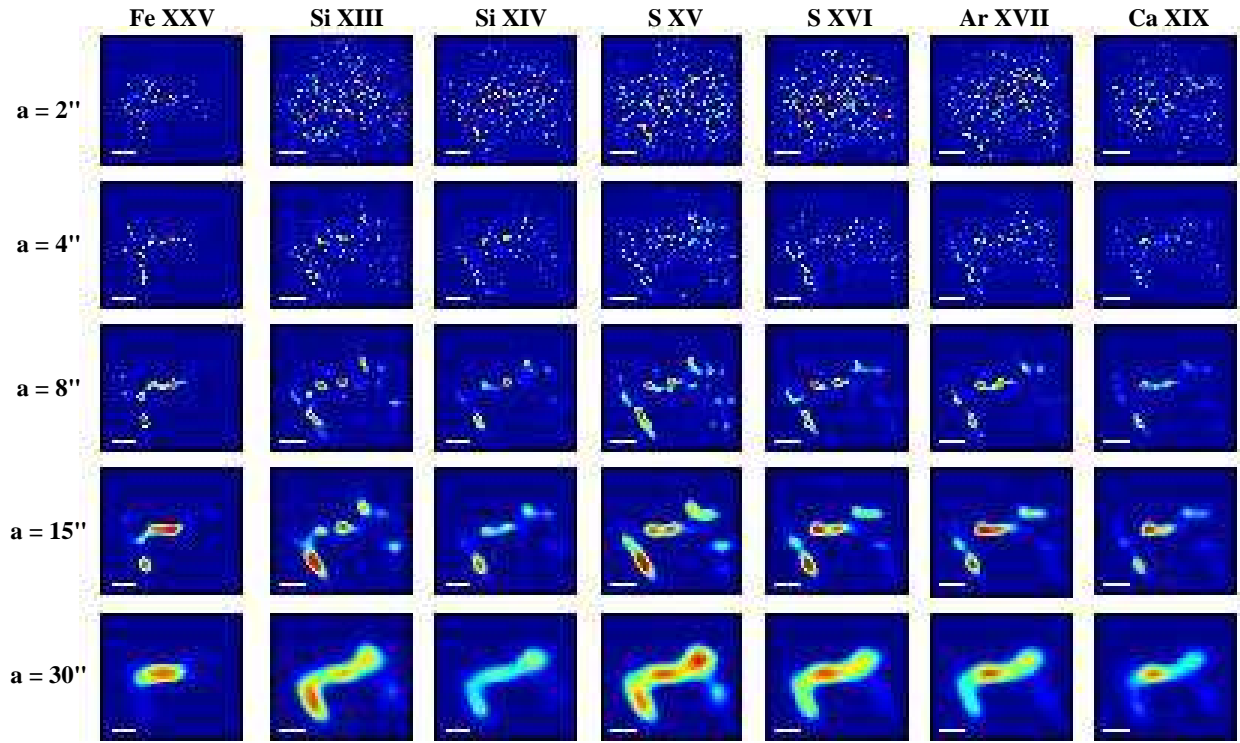}
\caption{Wavelet-transformed images of the seven ions detected in the {\it Chandra} ACIS observation of W49B for four different sized Mexican hats ($a$ = 1\arcsec, 2\arcsec, 8\arcsec, 15\arcsec, and 30\arcsec; recall 1 pixel = 0.492\arcsec). Images are normalized within each column, and the scale bar is \hbox{50\arcsec\ $\approx$ 1.93 pc} in size, for a distance $D$ = 8 kpc. Red and yellow signifies a lot of emission at those locations and scale, while blue denotes little-to-no power at those locations and scale. Raw images are continuum subtracted prior to the WTA (see text for details).} 
\label{fig:montage}
\end{figure}

We apply the discrete wavelet-transform analysis outlined in $\S$2.3 for the seven strong emission lines in the {\it Chandra} ACIS spectrum using Matlab Version 7.4. Figure~\ref{fig:montage} shows all the wavelet-transformed images for five different scales. As per the discussion in $\S$2.2, the fields display the X-ray power at the given sizes. At small scales, noise and random fluctuations dominate, and with increasing Mexican-hat sizes, the distribution and substructure of each ion becomes more evident. Consistent with the previous analysis, the Fe {\sc xxv} emission is absent from the Western region of the fields, whereas other ions are more symmetrically distributed. 

\begin{figure}
\epsscale{1.0}
\plotone{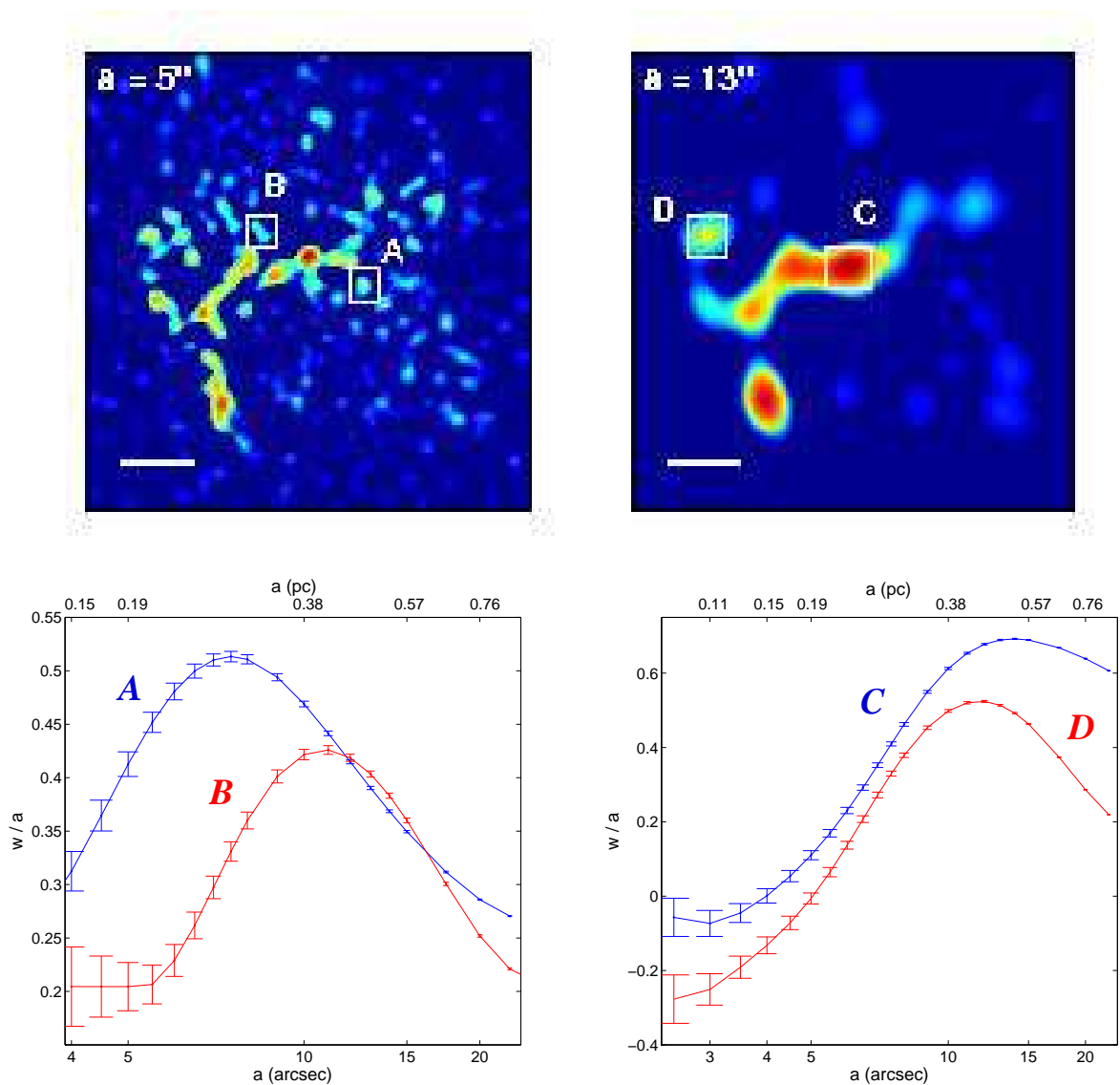}
\caption{{\it Above}: Two wavelet-transformed images ($w/a$) of Fe {\sc xxv} for Mexican-hat sizes $a$ = 5\arcsec and 13\arcsec. The scale bar is 50\arcsec\ $\approx$ 1.93 pc in size. White boxes identify the clumps with WTA plots underneath. {\it Below}: Example WTA plots ($w/a$ versus $a$) for individual clumps identified at each scale. The x-axis is given in units of arcseconds and parsecs. Errors are calculated via WTA assuming Poisson statistics. As in the synthetic data analysis, absolute maxima occur at the scale of most power, i.e., the size of an emitting region. Many Fe {\sc xxv} clump sizes are evident.} 
\label{fig:exampleclumps}
\end{figure}

\begin{figure}
\epsscale{0.9}
\plotone{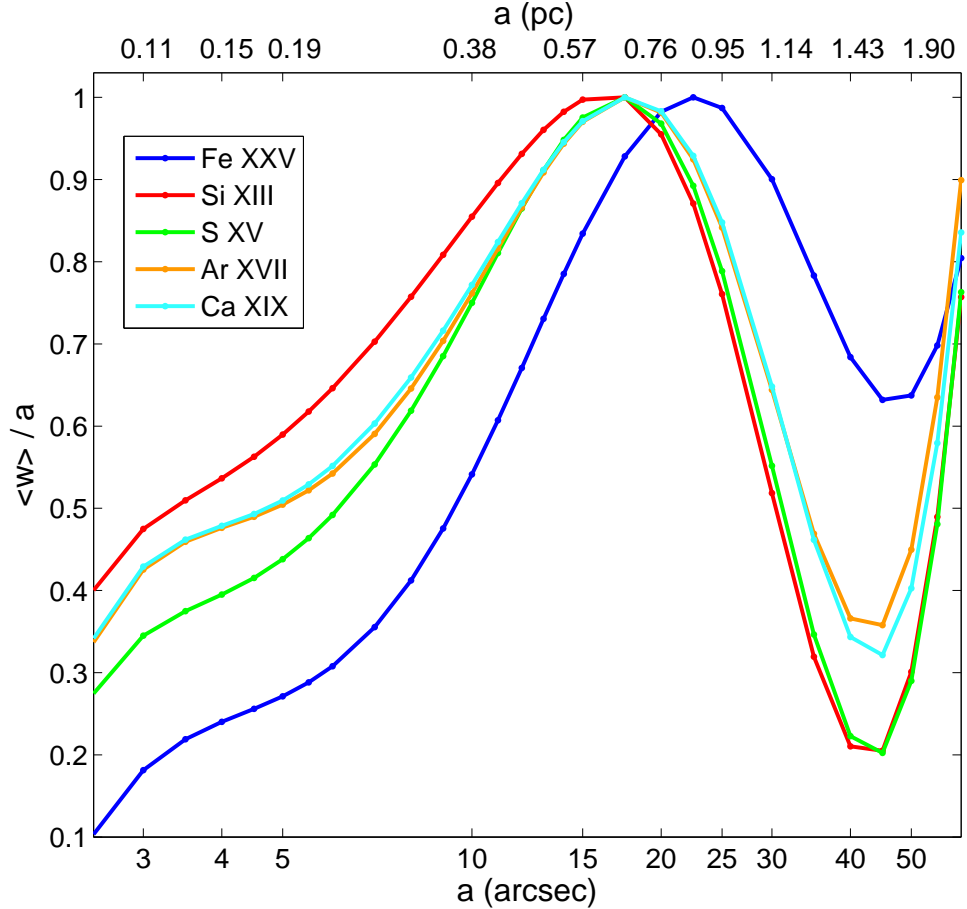}
\caption{The averaged WTA plot for the five strongest ions in the {\it Chandra} ACIS spectrum of W49B. Each point represents the power measured at the given scale for Fe {\sc xxv}, Si {\sc xiii}, S {\sc xv}, Ar {\sc xvii}, and Ca {\sc xix}. Curves are normalized to unity for easy visual comparison, and we ignored pixels with absolute maxima at the smallest scales to avoid contribution from noise. We estimate errors to be negligible ($\sim$ 10$^{-4}$) since each point represents the average value of $\approx$150,000 pixels. The absolute maxima are at $a_{{\rm max}}$ = 22.5\arcsec\ $\approx$ 0.86 pc for iron and $a_{{\rm max}}$ = 17.5\arcsec\ $\approx$ 0.67 pc for all other elements (where 0.492\arcsec\ = 1 pixel). Iron also contributes 40--60\% less power at scales $<$15\arcsec\ compared to the other ions. The turnover at very high $a$ values occurs as the Mexican-hat scale approaches the overall size of the remnant. We note that the $a_{{\rm max}}$ for all elements is much greater than the on- and off-axis point-spread function of ACIS, so we are not limited by instrumental resolution.} 
\label{fig:wa}
\end{figure}

As in the procedures outlined in $\S$2.4.2, we determine the scale of individual X-ray emitting regions as well as the intensity profile in our ACIS ion images. Figure~\ref{fig:exampleclumps} displays examples of identified clumps in the wavelet-transformed images of Fe {\sc xxv} and their WTA plots. The method distinguished successfully the location and scale of the substructure, and many different sizes of X-ray emitting regions are evident. Figure~\ref{fig:wa} shows the averaged WTA plot for the five strongest ions in the {\it Chandra} ACIS spectrum, Fe {\sc xxv}, Si {\sc xiii}, S {\sc xv}, Ar {\sc xvii}, and Ca {\sc xix}. We have ignored the global maxima at the smallest scales since these are due only to noise and have no physical interpretation. All the ions except Fe {\sc xxv} have the same global maximum, $a_{{\rm max}}$ = 17.5\arcsec\ = 0.67 pc. The Fe {\sc xxv} curve peaks at about 25\% larger value, with $a_{{\rm max}}$ = 22.5\arcsec\ = 0.86 pc and has 40--50\% more power than the other ions at scales of $\sim$30--60\arcsec. Thus, the characteristic sizes of X-ray emitting regions of the low-$Z$ ions are equal, whereas the scale of Fe {\sc xxv} clumps is greater. Fe {\sc xxv} contributes much less power at small scales relative to these ions as well: below scales of 15\arcsec, Fe {\sc xxv} emits 30--60\% less power than the other ions. These results indicate that the X-ray emitting plasma has a characteristic size, and this scale is similar for all ions except for Fe {\sc xxv} in W49B. We note that the $a_{\rm max}$ of all ions is sufficiently high to avoid being limited by noise. 

\begin{figure}
\epsscale{0.9}
\plotone{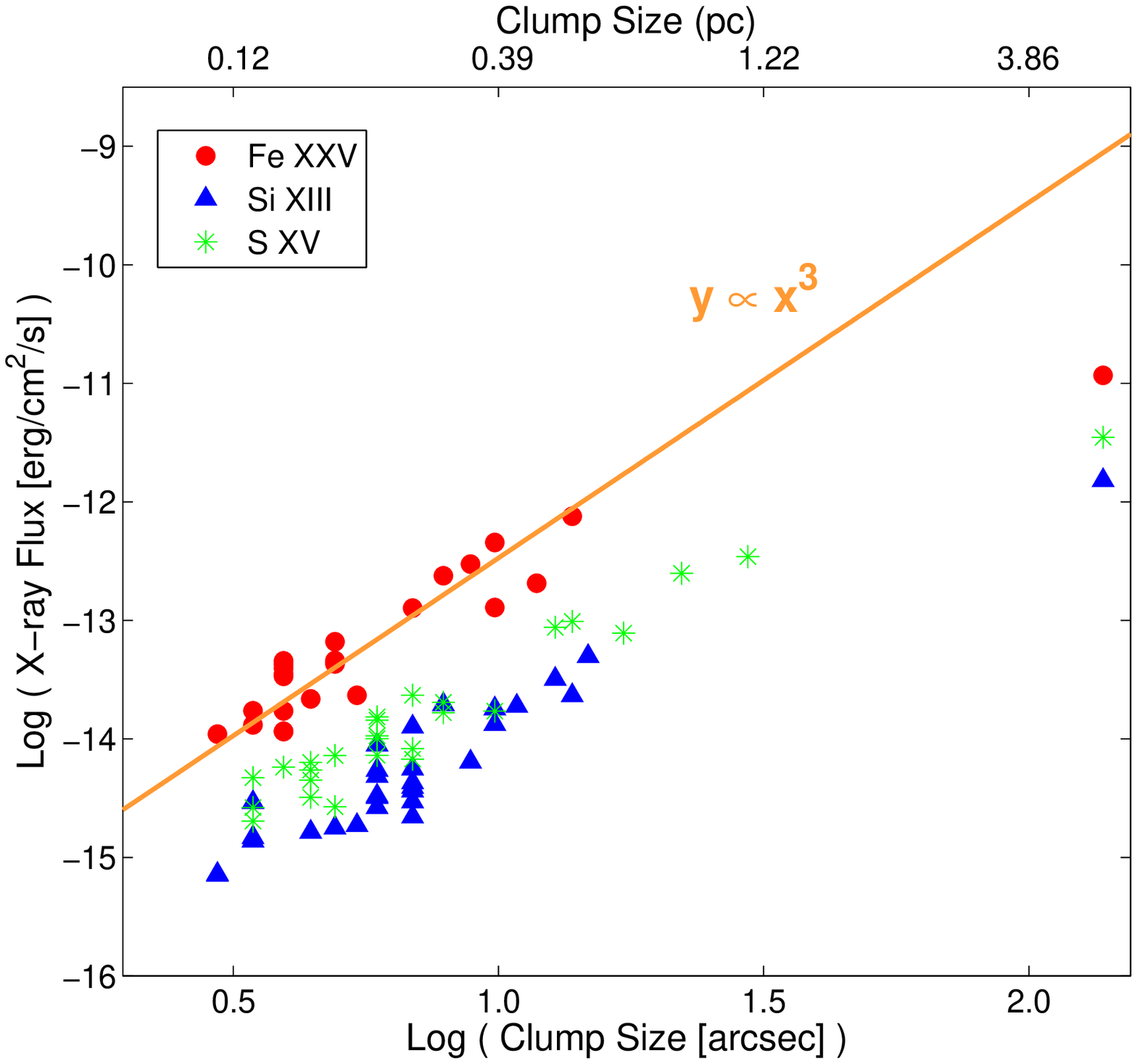}
\caption{X-ray flux versus clump size for Fe {\sc xxv} (red circles), Si {\sc xiii} (blue triangles), and S {\sc xv} (green stars). The points at the largest scale represent the flux and size of the entire remnant. The orange line represents the case if flux went as the cube of clump size (i.e., volume).} 
\label{fig:flux}
\end{figure}

We measure the counts from individual clumps using the {\sc ciao} command {\it dmextract} for a region of area equal to the clump size. We account for the background counts using a circular region with diameter 40\arcsec\ located 200\arcsec\ southwest of the remnant's center. We calculate the X-ray flux from individual clumps by multiplying the extracted counts by the centroid energy of its emission line and dividing by the duration of the observation. In Figure~\ref{fig:flux}, we plot the resulting clump sizes as a function of X-ray flux for Fe {\sc xxv}, Si {\sc xiii}, and S {\sc xv}. The points at the largest scale reflect the size and flux of these emission lines over the entire remnant. On the small scale, the X-ray fluxes go roughly as the cube of clump size (i.e., volume). However, when extrapolated to the full size of the remnant, such a relationship would severely overpredict the flux of the source. The line fluxes over the entire remnant are $\approx$65--100 times less than if they were proportional to volume. The inverse of these values gives the filling factor of the emission in the remnant: 1--1.5\%. This small value underscores the importance of considering substructure when using X-ray flux to estimate remnant properties (such as abundance ratios or elemental masses). 

\section{Discerning Heating and Explosion Properties}

From the above analyses, it is obvious that Fe {\sc xxv} has a distinct X-ray morphology. The distribution of Fe {\sc xxv} is more asymmetric than and disjoint from the other ions, and \hbox{X-ray} emitting regions of Fe {\sc xxv} are larger than those of the lighter ions. Two possible scenarios can account for these properties: the iron is insufficiently heated over large areas of the remnant, or it was ejected anisotropically during the supernova explosion. From the wavelet-transform analysis, we have identified quantitatively the regions with and without strong Fe {\sc xxv} emission, and we can model spectra in these locations to probe plasma properties and ultimately, to discern heating versus explosion effects. 

\subsection{Spectral Analysis \& Plasma Properties}

\begin{figure}
\epsscale{0.8}
\plotone{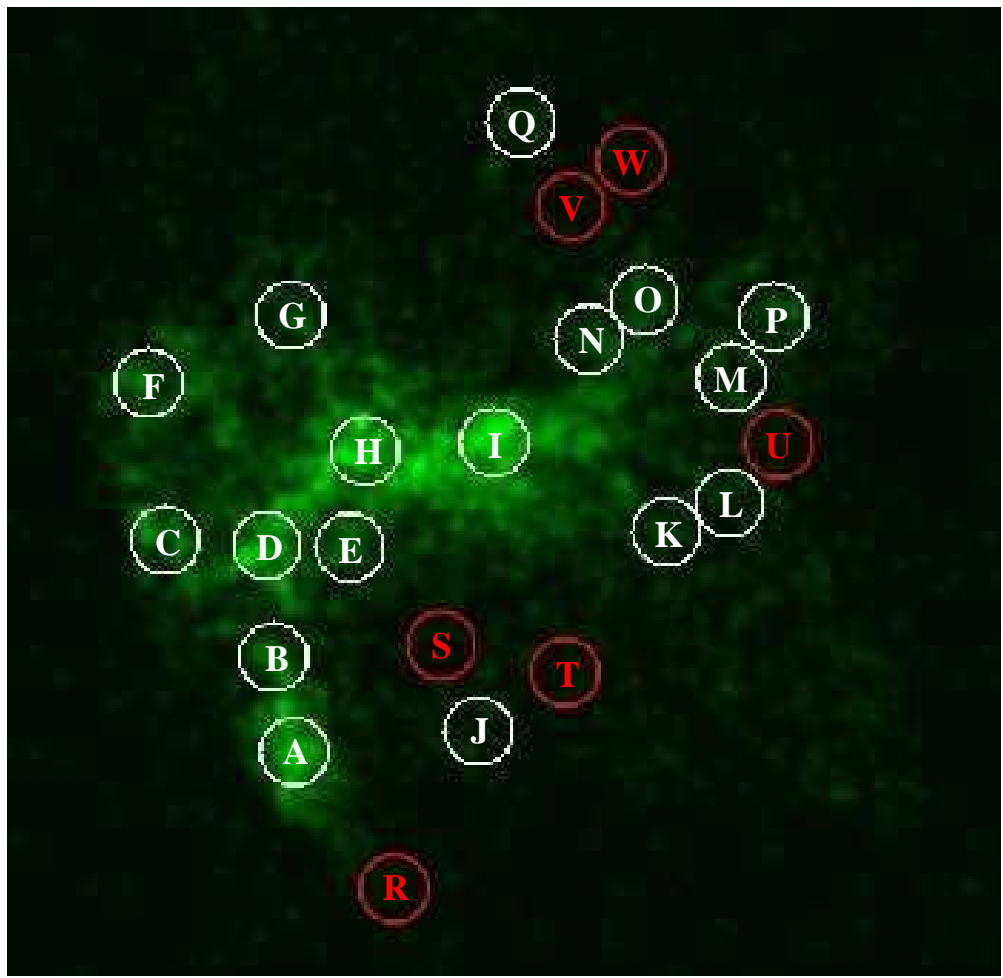}
\caption{Smoothed image of the iron emission in W49B. White circles identify regions with iron clumps (as identified by the wavelet-transform analysis) where we extracted \xmmn\ spectra. Red circles denote regions with weak iron emission where we extracted \xmmn\ spectra. Table 2 lists the best-fit electron temperatures and abundance ratios by mass relative to iron for each region.} 
\label{fig:locations}
\end{figure}

We extracted \xmmn\ spectra at seventeen locations identified to have Fe {\sc xxv} clumps (white circles A--Q in Figure~\ref{fig:locations}). W49B was observed three times by \xmm: 2004-Apr-03 for $\sim$20 ks (ObsID 0084100401); 2004-Apr-05 for $\sim$20 ks (ObsID 0084100501); and 2004-Apr-13 for 2.5 ks (\hbox{ObsID 0084100601}).  We downloaded reprocessed versions of the first two observations from the \xmmn\ Science Archive\footnote{http://xmm.esac.esa.int/xsa/}.  Both observations were taken with the PN and both MOS in the Full Window Imaging mode and with the Medium filter.  After filtering for intervals of high background flaring, the first observation was left with 14.5 ks of exposure for the PN and 18 ks for each MOS, and the second observation was left with 14.8 ks for the PN and 18 ks for each MOS.

For each of the locations circled in Figure~\ref{fig:locations}, we extracted spectra from the PN and both MOS detectors using the {\it evselect} tool. Our extraction regions were 10$''$ in radius. We selected events with PATTERN in the 0--4 range for the PN and 0--12 range for the MOS and with $\mathrm{FLAG}=0$ for both instruments. Response files were generated for each spectrum using the {\it rmfgen} and {\it arfgen} tools to take into account the selection criteria, extraction region size, and off-axis angle. Our background spectra were made from a source-free 250$''$ diameter circle located $\sim$10\arcmin\ northwest of the center of the remnant.

\begin{figure}
\epsscale{0.8}
\plotone{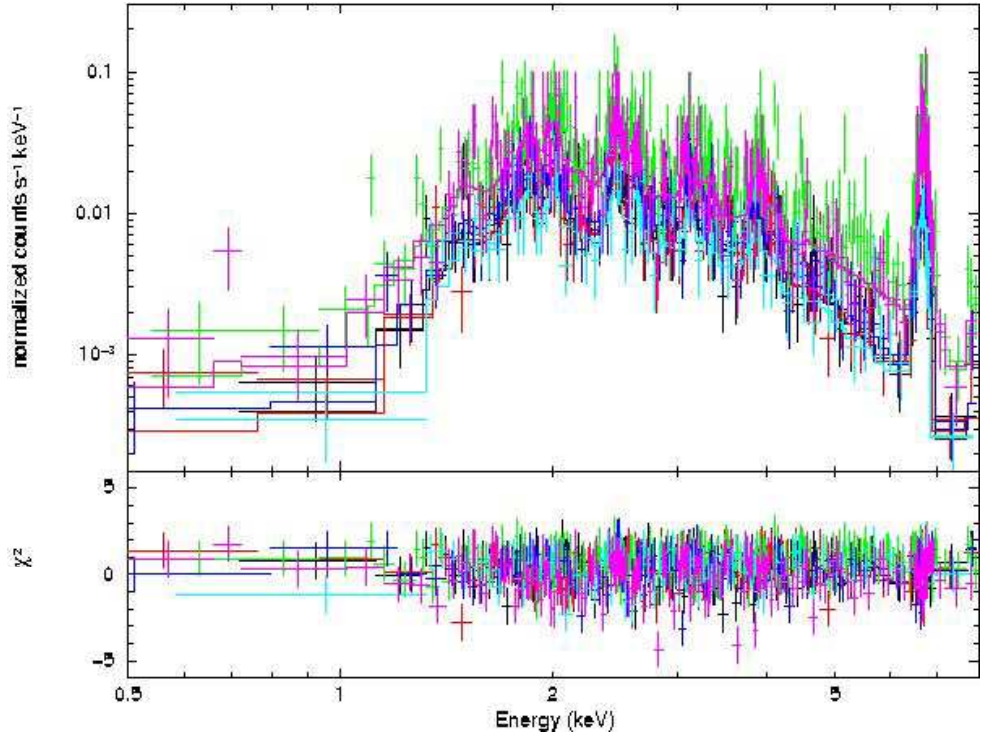}
\caption{Example \xmmn\ spectra, fits, and residuals for circle A in Figure 18. Best-fit electron temperatures and abundance ratios by mass relative to iron are given in Table 2. By fitting the six \xmm\ spectra from circle A simultaneously (each plotted in a different color), we improve our count statistics dramatically, and all spectra fits have reduced chi-squared values within 10\% of unity. For the model in this figure, $\chi^{2}$/d.o.f. = 1366/1410.} 
\label{fig:examplespectra}
\end{figure}

Spectra were fit using XSPEC Version 12.4.0. Data were grouped such that a minimum of five counts were in each energy bin, and the six MOS and PN spectra from each region were fit simulateneously to improve statistics. Standard weighting was used in calculating chi-squared values. We modeled the spectra with two components, one cool plasma with fixed solar abundances and one hotter plasma with varying supersolar abundances (similar to Miceli et al. 2006), in collisional ionization equilibrium (CIE) using the XSPEC model MEKAL (Mewe et al. 1985, 1986; Liedahl et al. 1995). Example spectra and fits from one region (circle A in Figure~\ref{fig:locations}) are plotted in Figure~\ref{fig:examplespectra}. CIE occurs when the electrons and ions reach equipartition, and the electrons are heated only by Coulumb collisions with the ions. The timescale $\tau$ to reach CIE is related to the electron density $n_{e}$ (in cm$^{-3}$) and the electron temperature $T_7$ (in units of $10^7$ K; Spitzer 1978): 

\begin{equation}
\tau = 7700 n_{e}^{-1} T_{7}^{3/2} {\rm years} . 
\label{eq:spitzer}
\end{equation}

\noindent
For a young supernova remnant like W49B, with an estimated age of $\sim$1000 years, CIE is only realistic with sufficiently high $n_{e}$.  From radio and infrared studies of W49B, $n_{e}$ is estimated to be quite high ($\sim$1000--8000 cm$^{-3}$; Keohane et al. 2007), making CIE plasmas a plausible scenario. In our fitting of the \xmm\ spectra, CIE models gave chi-squared values $\sim$10--15\% lower than non-equilibrium ionization (NEI) models. All seventeen produced good statistical descriptions of the data; Table 2 lists the chi-squared values and degrees of freedom associated with each model.

\begin{figure}
\epsscale{0.8}
\plotone{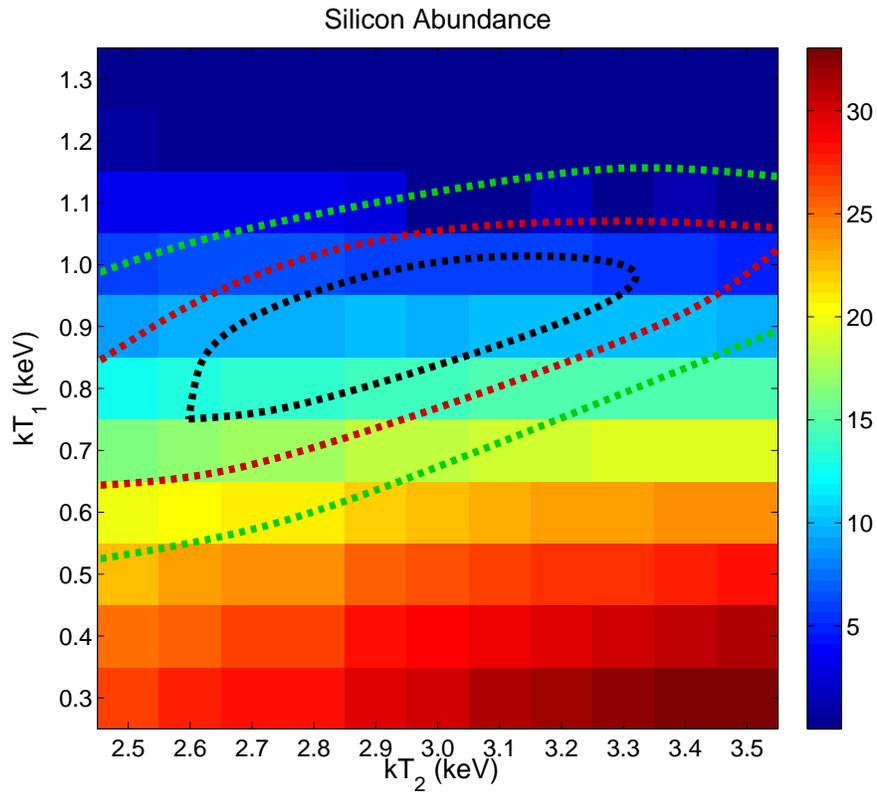}
\caption{Dependence of the silicon abundance on the temperature of the cool plasma ($kT_{1}$) and hot-plasma ($kT_{2}$) relative to solar. For constant $kT_1$, the abundance only changes by $\approx$10\%, while for constant $kT_{2}$, the abundance changes by a factor of 3. Based on these resulst, we conclude that the silicon emission arises mostly from the cool plasma.} 
\label{fig:si}
\end{figure}

The best-fit temperatures for the regions with iron emission are given in \hbox{Table 2}: the range in the cold plasma temperature is \hbox{$kT_{1} =$ 0.49--1.19 keV} (corresponding to \hbox{$T_{7} \sim$ 0.57--1.38}) and in the hot plasma temperature is \hbox{$kT_{2} =$ 1.81--3.68 keV} (corresponding to \hbox{$T_7 \sim$ 2.10--3.96}). Generally, the temperatures increase toward the Eastern side of the remnant. \cite{m06} found that the strong Si {\sc xiii} line arises from the cold plasma, while the other emission lines require the hot plasma. To test this result, Figure~\ref{fig:si} shows how the silicon abundance (denoted by color) changes as a function of the temperatures of the cool and hot components, $kT_1$ and $kT_2$. Confidence contours of $kT_1$ and $kT_2$ are overlaid to highlight their statistically viable values. For constant $kT_1$, the silicon abundance varies less than 10\%; by contrast, for constant $kT_2$, the silicon abundance changes by a factor of $\approx$3--4. Thus, we confirm the conclusions of \cite{m06} that the Si {\sc xiii} arises mostly from the cool plasma. 

\begin{figure}
\epsscale{0.8}
\plotone{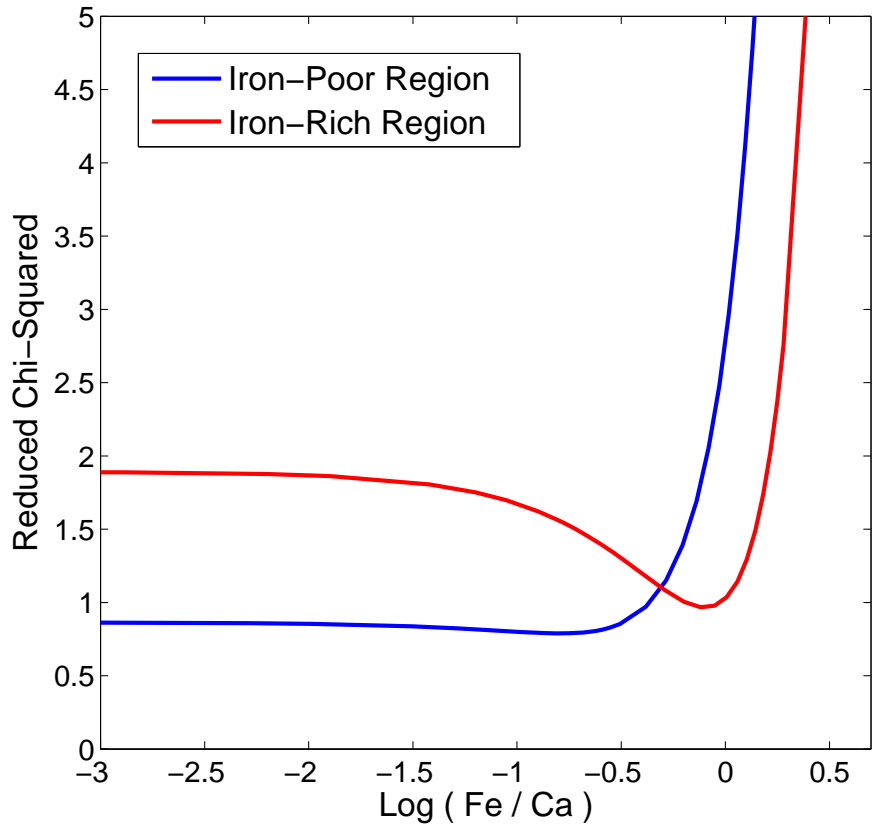}
\caption{Plot of reduced chi-squared as a function of iron abundance relative to calcium (with respect to solar by number). While the region with strong iron emission (circle A) has an absolute maximum at unity, the region with weak iron emission (circle V) has roughly constant chi-squared values for five orders then a sharp incline. This result statistically confirms the iron-depleted locations identified by WTA.} 
\label{fig:chisq}
\end{figure}

We extracted \xmmn\ spectra at six locations identified by WTA as weak in Fe {\sc xxv} emission and yet enhanced abundances of lighter ions (the red circles R--W in Figure~\ref{fig:locations}). We fit these spectra using a similar model as the regions with strong Fe {\sc xxv} emission: two CIE plasmas, one cool component with fixed solar abundances and one hot component with varying supersolar abundances. The resulting chi-squared values and degrees of freedom associated with each fit are given in Table 2; all six gave statistically good descriptions of the data. To verify the lack of Fe {\sc xxv} emission in these regions, we measured the reduced chi-squared of the fits as a function of iron abundance and compared the results to those from regions with strong iron emission. Figure~\ref{fig:chisq} plots the resulting reduced chi-squared values for iron abundances relative to calcium for one region with strong iron emission (circle A in Figure~\ref{fig:locations}) and for one region with weak iron emission (circle V in Figure~\ref{fig:locations}). Since Fe {\sc xxv} and Ca {\sc xix} form at the same electron temperatures, we plot the Fe/Ca ratio to establish whether iron is comparatively depleted. The iron-weak region has roughly constant reduced chi-squared values for iron-to-calcium abundances ranging $10^{-5}$ to 0.3 and rises sharply at larger ratios. By contrast, fits to the iron-strong emitting regions gave a local minimum in reduced chi-squared values at unity. Consequently, the results support the WTA identification of these regions as depleted in Fe {\sc xxv} emission. 

\begin{figure}
\epsscale{0.8}
\plotone{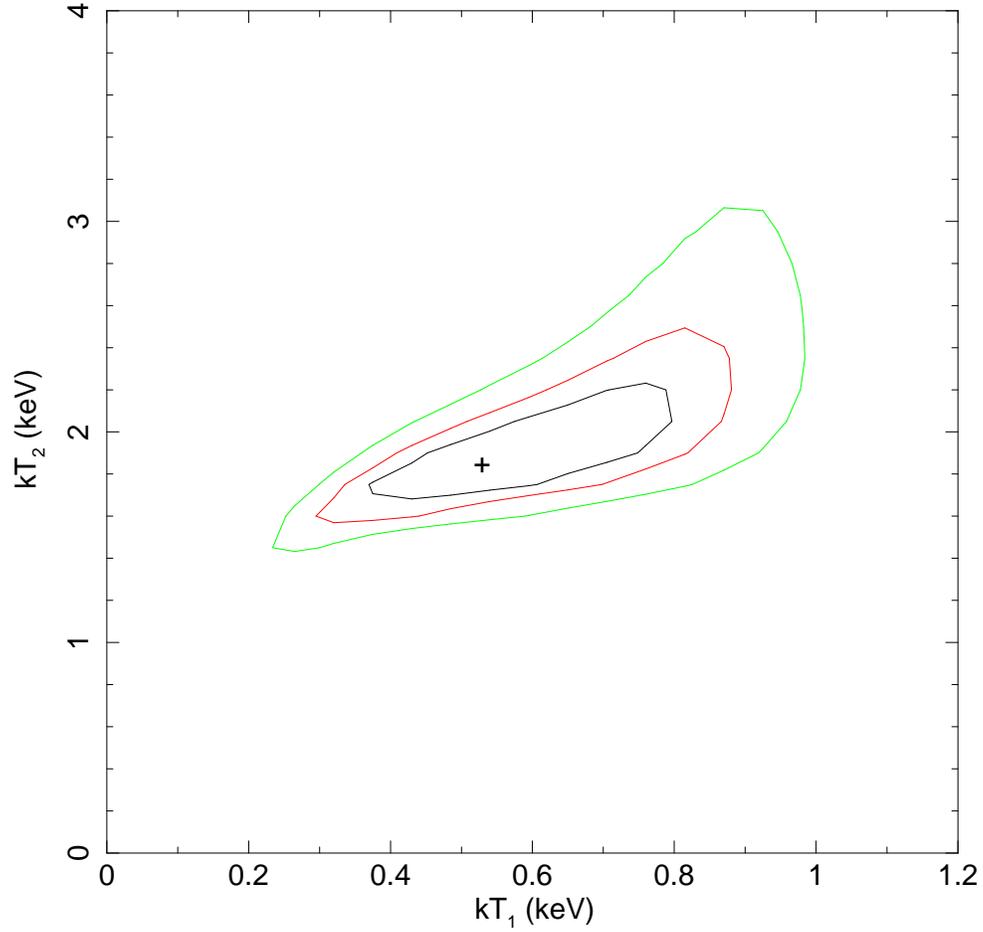}
\caption{68\%, 95\%, and 99\% confidence contours for $kT_{1}$ and $kT_{2}$ in a region with weak iron emission (as identified by our WTA analysis; circle V in Figure~\ref{fig:locations}). As the 99\% confidence region does not intersect with either axis, two plasmas with different temperatures are required throughout the remnant.} 
\label{fig:cc}
\end{figure}

To test whether two plasmas (and not one) are necessary in these areas with depleted iron emission, we varied the electron temperatures of the cool component ($kT_{1}$) and of the hot component ($kT_{2}$) over a grid of values while letting all other parameters float. This process produced the confidence contours plotted in Figure~\ref{fig:cc}. As the 99\% confidence region does not intersect the axes (i.e., electron temperatures of zero), plasmas with two temperatures are statistically necessary throughout the remnant. 

\begin{figure}
\epsscale{1.0}
\plotone{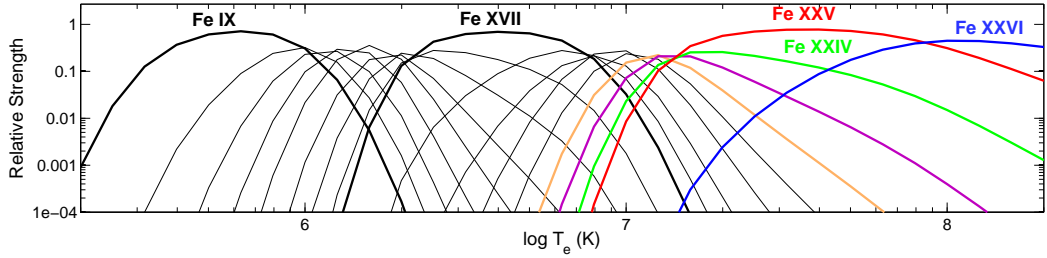}
\caption{Plot of relative strength of Fe ions as a function of electron temperature $T_{e}$ in a CIE plasma. The best-fit models of the regions with weak iron emission give electron temperatures corresponding to those where the Fe {\sc xxv} should have a prominent emission line. However, there is little-to-no Fe {\sc xxv} emission in these regions. As the plasma is sufficiently heated to irradiate iron, this result indicates that iron is greatly depleted in these regions. Thus, the anomalous distribution of iron must arise from an anisotropic ejection of nickel during the supernova explosion.} 
\label{fig:fe}
\end{figure}

We model the spectra with weak Fe {\sc xxv} emission with two CIE plasmas to estimate the electron temperatures there. The six regions gave best-fit hot plasma temperatures \hbox{$kT_{2} =$ 1.80--3.41 keV} (corresponding to \hbox{$T_{7} \sim$ 2.09--3.96}), very close to the values obtained for regions with strong Fe {\sc xxv} emission. At these electron temperatures, Fe {\sc xxv} should have a prominent line (see Figure \ref{fig:fe}) if iron is abundant there. However, little-to-no iron is X-ray illuminated in these regions, indicating that the iron is depleted from these parts of the remnant (with an abundance 1.49$\pm$0.88 relative to solar). Thus, insufficient heating is not a viable explanation for observed asymmetries, and the anomalous Fe {\sc xxv} distribution must arise from an anistropic ejection of iron. 

\subsection{Explosion Mechanism}

Since we have determined that the asymmetries in W49B result from an anisotropic ejection, we now explore supernova explosions that could lead to the observed Fe {\sc xxv} morphology. Bipolar explosions, a special class of core-collapse supernovae, are highly asymmetric explosions that would distribute heavy metals similarly to those in W49B. Core-collapse supernovae occur when the iron cores of massive stars ($M \sim 8-130 M_{\sun}$) collapse to form neutron stars or black holes. These explosions are categorized based on whether the progenitor lost its hydrogen envelope (Type Ib or Type Ic) or not (Type II). A core-collapse supernova could produce a bipolar explosion if the progenitor's core is undergoing sufficiently large rotation. Numerous mechanisms could be responsible for the resulting bipolar nature of the event \citep{rr02,mw99,mw01,dt92,w00,b07,bu07,kb07}.

Typically, core-collapse SNe have kinetic energies around $10^{51}$ ergs. However, in the last ten years, numerous detections of hypernovae --- supernovae with high ejecta velocities and sometimes large kinetic energies (e.g., Nomoto et al 2001) --- have challenged this paradigm. Some of these hypernovae may originate from bipolar explosions (Maeda \& Nomoto 2003). With the number of bipolar explosion/hypernovae candidates growing, the distinguishing properties of these extreme sources are being revealed. In the case of SN 1998bw, the [Fe {\sc ii}] lines were broader than the [O {\sc i}] $\lambda \lambda$ doublet, indicating that the $^{56}$Ni (which decays into iron) was ejected at larger velocities with respect to the oxygen \citep{m01}. A spherically symmetric model of the collapse could not account for this observation, and \cite{m02} successfully reproduced the result using an axisymmetric model. In this scenario, the kinetic energy is assumed to decrease as a function of solid angle from the polar axis, causing the nickel to be more efficiently synthesized and ejected at high velocity in that direction. Perpendicular to the rotational axis, mechanical heating decreases; thus, this phenomenon predicts that lower-$Z$ elements are ejected more isotropically and at smaller velocities than the nickel (e.g., Mazzali et al. 2005).

Generally, these asymmetries alter the chemical yield of bipolar explosions relative to typical CC SNe. While the typical core-collapse event ejects $\sim 0.07-0.15 M_{\sun}$ of $^{56}$Ni, bipolar explosions can have much greater nickel yields that also increase with asphericity, progenitor mass and explosion energy \citep{u08}. Examples of nickel masses derived for bipolar explosion candidates include $\sim$0.25--0.45 $M_{\sun}$ (2003dh: Mazzali et al. 2003), \hbox{$\sim$0.45--0.65 $M_{\sun}$} (2003lw: Mazzali et al. 2006), and $\sim$0.2--0.7 $M_{\sun}$ (1998bw: Woosley et al. 1999; Kaneko et al. 2007). 

The abundance ratios obtained in our spectral fits can be used to constain whether W49B may be the remnant of a core-collapse, bipolar explosion. \cite{m06} performed similar analyses for the bright, central region of W49B (circle I in Fig.~\ref{fig:locations}) and compared its abundance ratios to aspherical explosion models of extremely metal-poor stars by \cite{mn03}. Based on low silicon and sulfur abundances relative to iron, these authors exclude massive progenitors ($M_{{\rm ZAMS}} >$ 25 $M_{\sun}$) as well as large explosion energies ($E > 10^{52}$ erg). This conclusion assumes that the abundances in the central region are reflective of the entire remnant. However, the iron abundance is greatly enhanced in that location: as listed in Table 2, this area (circle I) has the lowest Si/Fe and S/Fe abundance ratios of all twenty-three regions in Figure~\ref{fig:locations}. Additionally, we have demonstrated in $\S$3 that the iron has statistically different symmetry, distribution, and substructure than the other elements. This attribute makes it imperative to average over many regions of the remnant to get an accurate representation of the abundance ratios. However, it is not adequate to estimate abundances from fits to the total spectrum because the central bar dominates the emission, giving uncharacteristicly low values for the lighter elements relative to iron. For example, the total spectrum gives a Si/Fe ratio of $\sim$0.18 and a S/Fe ratio is $\sim$0.07, close to the circle I values. We also note that the Miceli ratios are derived solely from the hot plasma, which affects the accuracy of the silicon abundance determination since it comes mostly from the cool plasma (as shown in Figure~\ref{fig:si}). Consequently, the Miceli estimate of the silicon abundance is low since it does not include the primary heating source for that element.

\begin{figure}
\epsscale{0.9}
\plotone{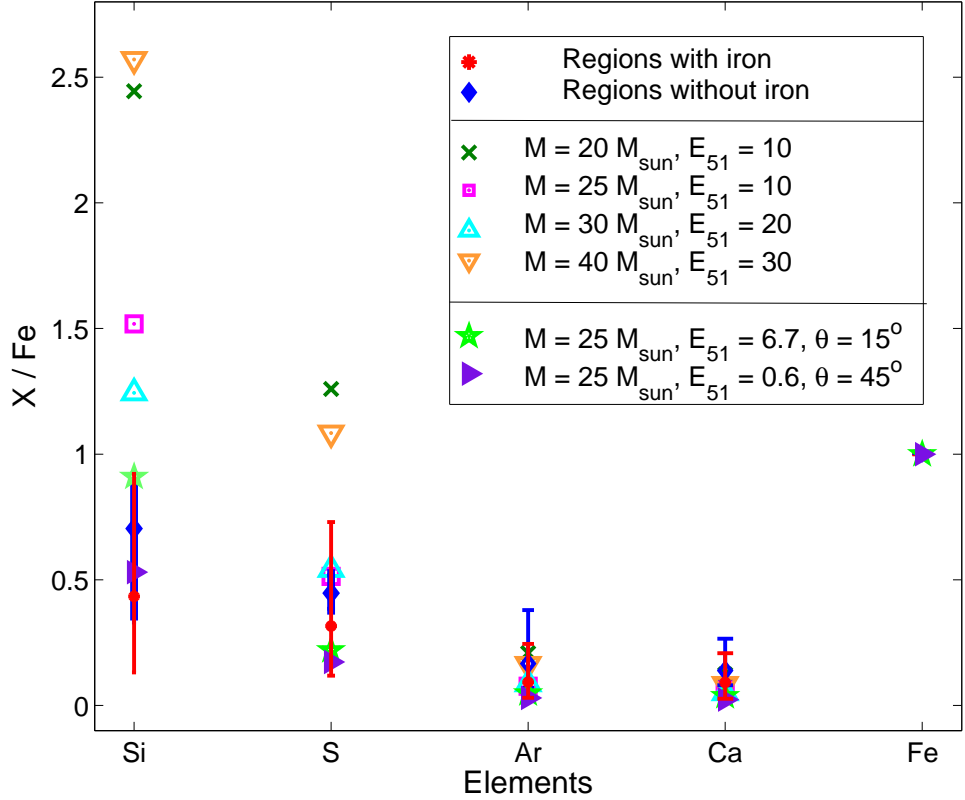}
\caption{Plot of mean abundances (by mass) and their range of values relative to iron of silicon, sulfur, argon and calcium. See Table 2 for the abundance ratios relative to iron for all twenty-three regions. For comparison, we show the abundance ratios predicted by models of four spherical explosions (from Nomoto et al. 2006) and two aspherical explosions (from Maeda \& Nomoto 2003) as well. The dispersion in abundance ratios, particularly silicon and sulfur, demonstrate the importance of averaging abundance ratios over the entire remnant. Generally, the spherical models are not consistent with the measured ratios, while the aspherical models succeed at fitting our obtained values, evidence that W49B is probably from a bipolar explosion.} 
\label{fig:abundances}
\end{figure}

To obtain an accurate measurement of the global abundance ratios in W49B, we fit the \xmm\ spectra from the twenty-three regions in Figure~\ref{fig:locations}, seventeen with strong Fe {\sc xxv} emission and six with depleted Fe {\sc xxv} emission. We plot the mean abundance ratios of the hot plasma (since the cool plasma is fixed to solar values) relative to iron for silicon, sulfur, argon, and calcium in Figure~\ref{fig:abundances} as well as list their values and 1-$\sigma$ errors in Table 2. The ratios have fairly large dispersions, with all spanning factors of 5--8. This result demonstrates that relative abundances of the elements change throughout the remnant, underscoring the importance of averaging over many regions in this calculation. We increase the silicon abundances plotted in Figure~\ref{fig:abundances} by 30\% from their Table 2 values to account for the contribution of the cool component. We estimated this factor by re-fitting the regions' spectra and fixing all elements of the cool plasma to solar metallicity except the silicon abundance, which was allowed to vary. Following this procedure, the combined silicon abundance of the two plasmas was 30\% greater than the hot component alone. 

In Figure~\ref{fig:abundances}, we compare our abundance ratios to those predicted by six core-collapse, Type Ic models: four energetic, spherical explosions \citep{n06} and two aspherical, bipolar explosions (25A, 25B: Maeda \& Nomoto 2003). The aspherical cases have the best agreement with our measured abundance ratios; both models' values are within the 1-$\sigma$ range for all elements in the W49B spectrum. The primary difference between the two aspherical models is their jet opening angles $\theta$: 25A has $\theta$ = 15$^o$ and 25B has $\theta$ = 45$^o$. For the fixed kinetic energy, a larger $\theta$ means less mechanical energy is released per unit solid angle in 25B, causing the explosion to appear less luminous and altering the nucleosynthetic products. Despite these distinctions, the dispersion in measured abundance ratios for W49B prevents us from discriminating between these two alternatives. Nonetheless, we can exclude the spherical explosion cases based on the silicon and sulfur ratios. Therefore, we find that a bipolar origin with typical or slightly larger kinetic energies of W49B is consistent with its global abundance ratios. 

\subsection{Mass Estimates}

The explosion mechanism can also be constrained by estimating the mass of metals in W49B. We perform this calculation using the elements' filling factor (as determined by the WTA) and the strength of the emission lines. Appendix B of \cite{l04} summarizes a similar method to find masses of individual regions of a source; we adapt this technique to the full W49B remnant and account for substructure using WTA. 

Masses are determined based on parameters given in the X-ray spectral fits. For a given electron temperature, the flux of an emission line is proportional to its emission measure, $EM_{{\rm line}} = \int n_{e} n_{i} dV$, where $n_{i}$ is the ion number density and $dV$ is the emitting volume. If $\zeta$ is defined as the ratio of electron and ion number densities, $n_{e}/n_{i}$, then $n_{i}$ is given by

\begin{equation}
n_i = \bigg( \frac{EM_{{\rm line}}}{\zeta V} \bigg)^{1/2}, 
\label{eq:ni}
\end{equation}

\noindent
where we have set $dV = V$, the volume of a chosen emitting region. From there, it is simple to derive the ion mass $M_{i}$ over the remnant:

\begin{equation}
M_{i} = f n_i V m_i = f m_i \bigg( \frac{EM_{{\rm line}} V}{\zeta}  \bigg)^{1/2}, 
\label{eq:clumpmass}
\end{equation}

\noindent
where $m_i$ is the mass per ion in grams and $f$ is the filling factor of the ion's emission over the remnant. 

The primary uncertainty in the above mass calculation is the electron number density $n_{e}$. However, since the plasma is in a collisional equilibrium state, we can set a physically motivated lower limit on $n_{e}$. By setting $\tau$ to the age of the remnant in Equation~\ref{eq:spitzer} \hbox{($\sim$ 1000 years)} and setting the electron temperature to the greatest value obtained by our spectral fits, \hbox{$T_{7} \sim$ 3.96}, we find \hbox{$n_e >$ 60 particles cm$^{-3}$}. Additionally, two reasonable scenarios define a strict range for $\zeta = n_{e}/n_{i}$ (see $\S$3 of Hughes et al. 2003 for a discussion). In the extreme case of a pure metal plasma, there are 50 electrons per Fe ion ($\zeta = 50$). Conversely, if the plasma has solar abundance (a large hydrogen content), there are 3$\times$10$^4$ electrons per Fe ion \hbox{($\zeta$ = 3$\times$10$^4$)}. An intermediate case, one where hydrogen mass is equal to that of metals (where \hbox{$\zeta$ = 600}), is also plausible. Our uncertainties are defined by the range of possible $n_{e}$, and our values are consistent with the results of \cite{k07}.

We can determine the filling factor $f$ using our WTA results and direct observables. The filling factor is defined as the ratio of the summed volume of individual clumps $V_{c}$ to the total volume of the entire source, $V_{T}$. Assuming the distance $D = 8$ kpc, we estimate the radius of X-ray emission in W49B is $R \approx$ 138\arcsec\ = 5.3 pc = \hbox{1.65 $\times 10^{19}$ cm}, giving \hbox{$V_{T} = \frac{4}{3} \pi R^{3} = 1.9 \times 10^{58}$ cm$^{3}$}. We find the summed volume of individual clumps by multiplying the mean volume of clumps from our WTA, $V_{{\rm mean}} = \frac{4}{3} \pi a_{{\rm max}}^{3}$, by the number of clumps, $N_{c}$. The number of clumps is determined simply by computing how many clumps with $V_{{\rm mean}}$ and a flux $F_{c}$ are necessary to account for the total flux $F_{T}$ of the source (as calculated using the procedure in $\S$3.3: 

\begin{equation}
N_{c} = \frac{F_{T}}{F_{c}} . 
\end{equation}

\noindent 
We opt to measure $N_{c}$ this way instead of counting all the clumps in the WTA images because the latter would require us to define a $w/a$ threshold for an individual clump. 

Putting all of these relations together, the filling factor $f$ is

\begin{equation}
f = \frac{V_{c}}{V_{T}} = \frac{N_{c} V_{{\rm mean}}}{V_{T}} = \frac{F_{T}}{F_{c}} \frac{a_{{\rm max}}^3}{R^3}. 
\label{eq:f}
\end{equation}

\noindent
Thus, 

\begin{equation}
M_{i} =  \frac{F_{T}}{F_{c}} a_{{\rm max}}^3 m_i \bigg( \frac{4 \pi EM_{{\rm line}}}{3 \zeta R^3}  \bigg)^{1/2}. 
\label{eq:all}
\end{equation}

We use Equation \ref{eq:all} to calculate the total mass of iron in W49B. From Figure~\ref{fig:wa}, \hbox{$a_{{\rm max}}$ $\approx$ 22.5\arcsec} = 0.86 pc for iron, and from Figure~\ref{fig:flux}, $N_{c} = F_T / F_c = 3.75$ for iron. These values give a filling factor $f$ = 0.016, an order of magnitude smaller than previous estimates that were based on assumptions about the extension of the emitting plasma \citep{m06}. Thus, a rigorous measurement of $f$ demonstrates that the value is actually much smaller, and iron especially must emit its power in concentrated locations of the remnant. 

We find the emission measure of iron $EM_{{\rm iron}}$ by the following procedure. Using the model from $\S$4.1, we fit the global Chandra {\it ACIS} X-ray spectrum of W49B (Figure~\ref{fig:spectrum}) to obtain the emission measure of the hot-plasma component: $EM_{{\rm hot}}$ = 1.805 $\times$ 10$^{59}$ cm$^{-3}$. To compute the emission measure of an individual line $EM_{{\rm line}}$, we reduce $EM_{{\rm hot}}$ by the ratio of flux in the line to that of the entire spectrum. X-ray fluxes are computed using the XSPEC command {\it flux} over the energy ranges of interest for a phenomenological model of the lines and continuum.Based on this procedure, iron contributes 23\% of the flux to the hot component, so \hbox{$EM_{{\rm iron}}$ = 4.29 $\times$ 10$^{58}$ cm$^{-3}$}. 

Using the equations above, we find an iron mass of $M_{{\rm iron}}$ $\approx$ 0.12 $M_{\sun}$ assuming a solar abundance plasma. For the case of hydrogen mass equal to that of metals, we obtain an iron mass of $M_{{\rm iron}}$ $\approx$ 0.80 $M_{\sun}$. The condition of a plasma in collisional equilibrium defines the upper-limit iron mass in W49B (since it gives the minimum $n_{e}$ necessary to reach CIE by the age of the remnant): $M_{{\rm iron}}$ $\approx$ 1.29 $M_{\sun}$. While this range of iron masses is large (reflecting the uncertainty in electron density), masses above 1.29 $M_{\sun}$ and below 0.12 $M_{\sun}$ can be excluded because it would require physically implausible metal abundances. Furthermore, the large iron mass cannot be attributed to a thermonuclear explosion since the abundance ratios relative to iron in Figure~\ref{fig:abundances} are much greater than those expected from Type Ia events. Although the derived iron masses are larger than those usually predicted for core-collapse supernovae, the lower limit is close to those values for typical bipolar explosions ($\approx 0.15 M_{\sun}$). Additionally, as noted in $\S$4.2, recent models of bipolar explosions demonstrate that nickel yields (which decays into iron) can actually be much greater than the typical core-collapse scenario (up to 4 $M_{\sun}$) with increased asphericity, explosion energy, or progenitor mass (Umeda \& Nomoto 2008). The anisotropic ejection of iron (as shown in $\S$4.1) and abundance ratios (given in $\S$4.2) in W49B are consistent with such an asymmetric explosion that would overproduce nickel. Generally, we emphasize that this calculation is an order-of-magnitude estimate, and precise measurement of the filling factor is an important step in obtaining a plausible range of masses. 

A few factors contribute to the uncertainty in our analyses. As noted previously, the largest source of error is the electron number density since we do not know precisely the plasma's composition. We attempt to limit this uncertainty by defining a plausible range in its value. Another source of error is the physical modeling of the X-ray spectrum. While the CIE model produces a chi-squared value lower than an NEI plasma, it is barely a statistically significant difference ($\sim$10\%). However, NEI models give abundance ratios within the errors of our CIE results, so we expect this uncertainty is minimal. Other sources of error are instrumental in nature: the off-axis point-spread function (PSF) effects with {\it Chandra} and the possible clump contamination with \xmmn\ from its limited spatial resolution. The PSF of {\it Chandra} ACIS increases as a function of energy and of off-axis angle. At its aimpoint, the PSF is $\sim$1\arcsec\ at 1.49 keV and $\sim$2\arcsec\ at 6.4 keV; at a distance 6\arcmin\ off-axis, the PSF increases to $\sim$3\arcsec\ at 1.49 keV and $\sim$5\arcsec\ at 6.4 keV (where we have characterized the PSF as the scale where the encircled energy fraction is 90\%). We anticipate that these effects do not alter our results much since the overall extent of W49B is small ($\sim$4\arcmin\ in diameter) and the characteristic clump sizes of all ions are many factors greater than these values. We minimize the uncertainty associated with \xmmn\ clump contamination by selecting emitting regions that were sufficiently isolated from other substructure. 

The SNR mass estimate in this paper is the first with the contribution of substructure rigorously defined. Measurement of ejecta masses requires detailed knowledge about the scale of the emitting regions since emission measure increases linearly with volume. Previous SNR studies assume generally that a source's emission is homogeneous (e.g., Hughes \& Singh 1994) or estimate the filling factor from basic arguments about the plasma's extension (e.g., Miceli et al. 2006) when calculating physical parameters. However, the complex ion distribution and low filling factor in W49B demonstrates the necessity for more rigorous approaches. Consequently, it is vital to probe accurately the role of substructure when determining global properties of a supernova remnant. 

\section{Summary}

In this paper, we have explored three methods that quantify the morphologies of extended sources. We have demonstrated the accuracy of these techniques on relevant synthetic data, and we have exemplified the utility of these tools by applying them to the Chandra {\it ACIS} observation of the SNR W49B.

Using a multipole power-ratio technique, we measured the asymmetry of the ion distributions in W49B. We discovered that the morphology of Fe {\sc xxv} is less elongated and more asymmetric than the lighter ions. We applied an adapted two-point correlation method called correlation-length analysis to compare the ions' spatial distributions. We determined that the Fe {\sc xxv} emission is more segregated and disjoint from that of the lower-$Z$ ions, which are well-mixed and similarly distributed throughout the remnant. Additionally, we measured properties of the X-ray ions' substructure in W49B using wavelet-transform analysis. We found that the mean clump scale of Fe {\sc xxv} is roughly 25\% larger than those of the lighter ions. Together, these results demonstrate quantitatively that Fe {\sc xxv} has a distinct X-ray morphology.

Two potential scenarios have been proposed to explain the anomalous properties of iron in W49B (Miceli et al. 2006; 2008; Keohane et al. 2007): insufficient heating of iron over large areas of the remnant or anisotropic ejection of iron during the supernova explosion. By fitting \xmmn\ spectra in twenty-three locations with strong and weak Fe {\sc xxv} emission (as identified by our wavelet-transform analysis), we determined that the electron temperatures throughout the remnant should heat any iron there to radiate at X-ray temperatures. Thus, iron is strongly depleted from the identified regions, and the distinct X-ray morphology must arise from an anisotropic ejection. Therefore, any models used to describe the explosion mechanism of W49B must include a means for the iron to be ejected an asymmetric way. 

We further constrained the nature of the explosion by measuring the mean abundance ratios of elements in our twenty-three X-ray spectral fits. The ratios varied tremendously in different locations, demonstrating the importance of averaging the ratios over many regions to obtain the global abundances in W49B. We found that the mean ratios are consistent with bipolar explosion models of massive stars. This type of explosion could also account for the anistropic ejection of nickel since heavy elements are preferentially ejected along the polar axis of the progenitor in this scenario. We calculated the total iron mass in W49B by determining the filling factor (using our wavelet-transform analysis) and the emission measure. We found that the iron mass and abundance ratios are broadly consistent with a core-collapse supernova. 

This work is a first attempt to rigorously describe the physical properties of supernova remnants. SNRs play a vital role in our galaxy, and the wealth of data available on these sources is tremendous. Robust methods for comparison within and between these complex targets can advance SNR science as well as probe many poorly understood phenomena. We plan to apply the techniques presented here to other sources to constrain further processes as particle acceleration, interactions with environments, and explosion mechanisms. 

\acknowledgements

We thank K. C. Schlaufman and S. E. Woosley for helpful discussions and contributions. This work is supported by DOE SciDAC DE-FC02-01ER41176 (LAL and ER-R) and a National Science Foundation Graduate Research Fellowship (LAL).

\clearpage

%%%% TABLE 1: Power Ratios
\begin{deluxetable}{cccc}
\tablecolumns{4}
\tablewidth{0pc}
\tabletypesize{\footnotesize}
\tablecaption{Power Ratios and Associated 90\% Confidence Limits}
\tablehead{\colhead{} & \colhead{$P_{2}/P_{0}$} & \colhead{$P_{3}/P_{0}$} & \colhead{$P_{4}/P_{0}$} \\
          \colhead{Ion} & \colhead{($\times$ 10$^{-7}$)} & \colhead{($\times$ 10$^{-7}$)} & \colhead{($\times$ 10$^{-7}$)} }
\startdata
Fe {\sc xxv} & 72.68$^{+16.95}_{-15.28}$ & 15.88$^{+2.46}_{-3.46}$ & 0.38$^{+0.48}_{-0.22}$ \\
Si {\sc xiii} & 137.54$^{+17.47}_{-16.24}$ &  2.87$^{+1.18}_{-1.44}$ & 0.92$^{+0.48}_{-0.37}$  \\
S {\sc xv} & 141.64$^{+13.94}_{-12.82}$ & 4.15$^{+1.09}_{-1.38}$   & 1.45$^{+0.55}_{-0.41}$  \\
Ar {\sc xvii} & 141.29$^{+23.36}_{-19.54}$ &  3.45$^{+1.34}_{-1.56}$  & 1.51$^{+0.73}_{-0.79}$ \\
Ca {\sc xix} & 136.19$^{+27.07}_{-13.81}$ & 3.42$^{+1.82}_{-1.66}$   & 1.76$^{+0.83}_{-0.65}$ \\
\enddata
\end{deluxetable}

%%%% TABLE 2: Abundances 

\begin{deluxetable}{cccccccc}
\tablecolumns{8}
\tablewidth{0pc}
\tabletypesize{\scriptsize}
\tablecaption{Best-Fit Temperatures and Hot-Plasma Abundance Ratios (Relative to Iron by Mass)}
\tablehead{\colhead{Region} & \colhead{$kT_{1}$} & \colhead{$kT_{2}$} & \colhead{Si/Fe} & \colhead{S/Fe} & \colhead{Ar/Fe} & \colhead{Ca/Fe} & \colhead{$\chi^{2}$/d.o.f.} \\
     \colhead{} & \colhead{(keV)} & \colhead{(keV)}      &   & & & & }       
\startdata
\cutinhead{Regions with Strong Iron Emission}
A & 0.91$\pm$0.10 & 2.92$\pm$0.31 & 0.46$\pm$0.14 & 0.26$\pm$0.08 & 0.08$\pm$0.03 & 0.08$\pm$0.03 & 939.7 / 918 \\
B & 0.59$\pm$0.11 & 1.99$\pm$0.12 & 0.31$\pm$0.06 & 0.19$\pm$0.03 & 0.05$\pm$0.01 & 0.05$\pm$0.01 & 482.5 / 514 \\
C & 1.05$\pm$0.11 & 3.33$\pm$0.33 & 0.57$\pm$0.22 & 0.56$\pm$0.22 & 0.12$\pm$0.05 & 0.14$\pm$0.06 & 694.3 / 701 \\
D & 0.70$\pm$0.10 & 2.51$\pm$0.17 & 0.18$\pm$0.06 & 0.12$\pm$0.03 & 0.04$\pm$0.01 & 0.03$\pm$0.01 & 877.9 / 934 \\
E & 0.67$\pm$0.17 & 2.43$\pm$0.17 & 0.40$\pm$0.10 & 0.32$\pm$0.07 & 0.08$\pm$0.02 & 0.07$\pm$0.02 & 832.7 / 855 \\
F & 1.19$\pm$0.15 & 3.68$\pm$0.38 & 0.33$\pm$0.21 & 0.17$\pm$0.08 & 0.09$\pm$0.04 & 0.08$\pm$0.04 & 461.5 / 480 \\
G & 0.49$\pm$0.14 & 2.46$\pm$0.23 & 0.27$\pm$0.09 & 0.14$\pm$0.04 & 0.05$\pm$0.02 & 0.04$\pm$0.01 & 387.9 / 416 \\
H & 0.79$\pm$0.13 & 2.55$\pm$0.13 & 0.30$\pm$0.06 & 0.23$\pm$0.04 & 0.08$\pm$0.02 & 0.06$\pm$0.01 & 1358 / 1405 \\
I & 0.80$\pm$0.13 & 2.54$\pm$0.13 & 0.13$\pm$0.04 & 0.14$\pm$0.02 & 0.04$\pm$0.01 & 0.04$\pm$0.01 & 1366 / 1410 \\
J & 0.66$\pm$0.29 & 1.81$\pm$0.32 & 0.28$\pm$0.17 & 0.15$\pm$0.07 & 0.08$\pm$0.04 & 0.06$\pm$0.03 & 351.3 / 388 \\
K & 0.60$\pm$0.34 & 1.86$\pm$0.12 & 0.39$\pm$0.08 & 0.30$\pm$0.05 & 0.07$\pm$0.02 & 0.07$\pm$0.02 & 663.9 / 676 \\ 
L & 0.95$\pm$0.11 & 3.02$\pm$0.34 & 0.33$\pm$0.14 & 0.50$\pm$0.21 & 0.25$\pm$0.11 & 0.21$\pm$0.09 & 617.6 / 657 \\
M & 0.97$\pm$0.11 & 2.18$\pm$0.29 & 0.77$\pm$0.46 & 0.51$\pm$0.31 & 0.17$\pm$0.10 & 0.14$\pm$0.08 & 644.7 / 738 \\
N & 0.91$\pm$0.21 & 2.30$\pm$0.22 & 0.39$\pm$0.10 & 0.29$\pm$0.09 & 0.07$\pm$0.03 & 0.06$\pm$0.02 & 819.3 / 868 \\
O & 0.92$\pm$0.28 & 1.93$\pm$0.18 & 0.44$\pm$0.12 & 0.23$\pm$0.07 & 0.07$\pm$0.02 & 0.06$\pm$0.02 & 863.9 / 941 \\
P & 1.00$\pm$0.13 & 2.42$\pm$0.32 & 0.70$\pm$0.33 & 0.50$\pm$0.26 & 0.11$\pm$0.06 & 0.12$\pm$0.06 & 820.0 / 818 \\
Q & 0.69$\pm$0.11 & 2.28$\pm$0.53 & 0.31$\pm$0.13 & 0.23$\pm$0.10 & 0.03$\pm$0.01 & 0.08$\pm$0.03 & 207.9 / 245 \\
\cutinhead{Regions with Weak Iron Emission}
R & 0.79$\pm$0.11 & 3.41$\pm$1.03 & 0.56$\pm$0.23 & 0.47$\pm$0.19 & 0.38$\pm$0.15 & 0.22$\pm$0.09 & 217.0 / 271 \\
S & 0.89$\pm$0.25 & 2.17$\pm$0.45 & 0.53$\pm$0.37 & 0.37$\pm$0.25 & 0.07$\pm$0.05 & 0.08$\pm$0.06 & 302.0 / 339 \\
T & 0.50$\pm$0.16 & 1.80$\pm$0.19 & 0.55$\pm$0.24 & 0.53$\pm$0.17 & 0.12$\pm$0.05 & 0.13$\pm$0.04 & 327.7 / 376 \\
U & 1.02$\pm$0.07 & 2.40$\pm$0.37 & 0.26$\pm$0.23 & 0.49$\pm$0.44 & 0.23$\pm$0.20 & 0.26$\pm$0.23 & 558.1 / 627 \\
V & 0.55$\pm$0.14 & 1.85$\pm$0.23 & 0.65$\pm$0.28 & 0.40$\pm$0.15 & 0.11$\pm$0.05 & 0.08$\pm$0.04 & 276.9 / 319 \\
W & 0.83$\pm$0.15 & 2.16$\pm$1.11 & 0.70$\pm$0.52 & 0.42$\pm$0.31 & 0.09$\pm$0.07 & 0.06$\pm$0.04 & 242.5 / 283 \\
\enddata
\end{deluxetable}

\clearpage

\clearpage

\clearpage

\clearpage

\end{document}